\documentclass{aa}
\usepackage{graphicx}
\usepackage[authoryear]{natbib}
\bibpunct[]{(}{)}{;}{a}{}{,}

\begin{document}

\title{Warm dust and aromatic bands as quantitative probes of star-formation activity
     \thanks{Based on observations with ISO, an ESA project
               with instruments funded by ESA member states (especially
               the PI countries: France, Germany, the Netherlands, and
               the United Kingdom), and with participation of ISAS and NASA.}
}
\titlerunning{Warm dust, PAHs and star formation}

\author{N. M. F\"orster Schreiber\inst{1}, H. Roussel\inst{2}, 
        M. Sauvage\inst{3}, V. Charmandaris\inst{4}}

\authorrunning{F\"orster Schreiber, Roussel et al.}

\offprints{H. Roussel}

\institute{Leiden Observatory, Leiden University, Postbus 9513,
           RA Leiden, The Netherlands
\and
California Institute of Technology, Pasadena, CA 91125, USA
\and
CEA/DSM/DAPNIA/Service d'Astrophysique, C. E. Saclay,
    F-91191 Gif sur Yvette CEDEX, France 
\and
Cornell University, Astronomy Department, Ithaca, NY 14853, USA \\
\email{forster@strw.leidenuniv.nl, hroussel@irastro.caltech.edu, msauvage@cea.fr, \\
       \hspace*{0.9cm} vassilis@astro.cornell.edu}
}

\date{Received 2 January 2004; accepted 2 February 2004}

\abstract{
We combine samples of spiral galaxies and starburst systems observed
with ISOCAM on board ISO to investigate the reliability of mid-infrared
dust emission as a quantitative tracer of star formation activity.
The total sample covers very diverse galactic environments and
probes a much wider dynamic range in star formation rate density
than previous similar studies. We find that both the
monochromatic 15\,$\mu$m continuum and the $5 - 8.5\,\mu$m emission
constitute excellent indicators of the star formation rate as quantified
by the Lyman continuum luminosity $L_\mathrm{Lyc}$, within specified
validity limits which are different for the two tracers.
Normalized to projected surface area, the 15\,$\mu$m continuum
luminosity $\Sigma_\mathrm{15\,\mu m,ct}$ is directly proportional to 
$\Sigma_\mathrm{Lyc}$ over several orders of magnitude.
Two regimes are distinguished from the relative offsets in the observed
relationship: the proportionality factor increases by a factor of
$\approx 5$ between quiescent disks in spiral galaxies, and moderate
to extreme star-forming environments in circumnuclear regions of spirals
and in starburst systems.  The transition occurs near
$\Sigma_\mathrm{Lyc} \sim 10^{2}~\mathrm{L_{\odot}\,pc^{-2}}$ and is
interpreted as due to very small dust grains starting to dominate
the emission at 15\,$\mu$m over aromatic species above this threshold.
The $5 - 8.5\,\mu$m luminosity per unit projected area is also
directly proportional to the Lyman continuum luminosity, with a single
conversion factor from the most quiescent objects included in the sample
up to $\Sigma_\mathrm{Lyc} \sim 10^{4}~\mathrm{L_{\odot}\,pc^{-2}}$, where
the relationship then flattens.
The turnover is attributed to depletion of aromatic band carriers
in the harsher conditions prevailing in extreme starburst environments.
The observed relationships provide empirical calibrations useful for estimating
star formation rates from mid-infrared observations, much less affected by
extinction than optical and near-infrared tracers
in deeply embedded \ion{H}{ii} regions and obscured starbursts,
as well as for theoretical predictions from evolutionary synthesis models.

}

\maketitle

\section{Introduction}  \label{Sect-intro}

Star formation is a fundamental process of galaxy formation and evolution.
Estimates of the star formation rate (SFR) in galaxies at all redshifts are key
indicators of the efficiency and mechanical feedback effects of star formation
activity, of the chemical evolution of the interstellar and intergalactic
medium, and, ultimately, of the cosmic star formation history.

Commonly used probes of the SFR include
photospheric emission from hot stars in the ultraviolet, nebular
H and He recombination lines as well as fine-structure lines arising in
\ion{H}{ii} regions from optical to radio wavelengths, and the total infrared
luminosity ($\lambda = 8 - 1000\,\mu$m), the bulk of which
is due to heated dust reprocessing the interstellar radiation field
\citep[see, e.g., the review by][]{Ken98}.  However, ultraviolet, optical,
and even near-infrared diagnostics are subject to potentially large 
uncertainties because of extinction in deeply embedded young star-forming sites
and in nuclear regions of galaxies.  Nebular lines may be difficult to
measure when intrinsically weak or superposed over a strong continuum. 
While dust emission suffers very little from extinction effects
and is usually strong in star-forming environments, the total infrared
luminosity is difficult to evaluate because it is generally derived from
observations in a few wavelength intervals which do not constrain the spectral
energy distribution accurately.  Moreover, a cirrus-like component unrelated
to star-forming regions can contribute substantially to the far-infrared
output of galaxies \citep{Hel86, Sau92}.

Mid-infrared emission (MIR, $\lambda = 5 - 20\,\mu$m) provides an
alternative probe of star formation activity.  The ``classical'' spectrum of
star-forming sources exhibits broad emission features often referred to as
``unidentified infrared bands'' (UIBs) and of which the most prominent dominate
the $6 - 13\,\mu$m range, and a continuum rising towards
long wavelengths at $\lambda \ga 11\,\mu$m 
\citep[see the reviews by][]{Geb97, Tok97, Ces99, Gen00}.
Various carbonaceous materials have been proposed to carry the UIBs, including the
popular polycyclic aromatic hydrocarbons (PAHs; e.g. \citealt{Leg84}) that
we adopt hereafter.
\citet{Peeters02} have analysed their shape and relative amplitude
variations in different classes of Galactic objects.
The continuum emission is generally attributed to very small dust grains
(VSGs; e.g. \citealt{Des90}) about which little is known.  Superposed on these
PAH and VSG components, H recombination lines and fine-structure lines of various
metals originating in \ion{H}{ii} and photodissociation
regions are observed as well
\citep[e.g.][]{Stu00}.  These lines may however not always be measurable
because of their weakness or of insufficient spectral resolution.

Numerous past studies have established that PAH
and $\lambda \ga 11\,\mu$m
continuum emission trace well star-forming regions but their usefulness as
{\em quantitative\/} diagnostics is still debated.  Complications arise from
the different nature of the emitting particles and by their being out of
thermal equilibrium under most radiation field conditions, undergoing large
temperature fluctuations of several 100~K
\citep[e.g.][]{Greenberg74, Draine85, Puget89}.
In addition, although both species are predominantly heated by energetic
radiation, PAHs can also be excited by softer optical and near-ultraviolet
photons as indicated by their detection in the diffuse interstellar medium,
in regions of insufficient far-ultraviolet energy density to account for their
heating \citep[e.g.][]{Sel90, Mat96, Uch98, Uch00, Li02}.
Furthermore, empirical evidence indicates that the $\lambda \ga 11\,\mathrm{\mu m}$
emission is produced by a mixture of dust particles akin to PAHs (or at
least whose flux variations follow well those of PAHs) and of VSGs
\citep[e.g.][]{Hony01, Rou01b}. The first component
is best seen in quiescent environments such as disks of spiral galaxies while
the second becomes prominent in active star formation sites.  It is not yet
clear how their combined emission varies over a large dynamic range in
star formation intensity.

On the other hand, spatially resolved studies of Galactic and Magellanic
Clouds \ion{H}{ii} regions have revealed that both the PAH features and
the VSG continuum are produced in the vicinity of massive stars, the 
former arising in photodissociation regions (PDRs) at the interface between
ionized and molecular gas and the latter peaking closer to the ionizing stars
\citep[e.g.][]{Geb97, Tok97, Ver96, Cre99, Con00}.
MIR imaging of external galaxies has shown that bright emission from
both components is closely associated with active star-forming sites
on large scales as well \citep[e.g.][]{Mir98, Mat99, Rou01c, FS03}.
A strong coupling with the SFR may thus exist and has been demonstrated
for disks of spiral galaxies by \citet{Rou01c}.  Specifically, these
authors found that the broadband $5 - 8.5\,\mu$m and
$12 - 18\,\mu$m fluxes vary linearly with the H$\alpha$ line
flux in the disks of 44 spirals.
\citet{FS03} also found a direct proportionality between the monochromatic
15\,$\mu$m continuum ($\Delta\lambda = 0.4\,\mu$m) and the
[\ion{Ar}{ii}] 6.99\,$\mu$m line emission in the nearby starbursts
M\,82, NGC\,253, and NGC\,1808, down to spatial scales of $\sim 100$~pc.

In this paper, we pursue the work of \citet{Rou01c} and \citet{FS03} by
combining samples of spiral and starburst galaxies observed
with the ISOCAM instrument \citep{Ces96} on board the
Infrared Satellite Observatory \citep[ISO;][]{Kes96}.
The merged sample covers diverse environments ranging from quiescent
galactic disks to infrared-luminous merging systems.  This
allows us to extend the investigation to higher activity levels
and to test whether previous results restricted to specific environments can
be generalized into more universal relationships.  We derive the dependence
of the PAH-dominated $5 - 8.5\,\mu$m emission and the VSG-probing
monochromatic $15\,\mu$m continuum on the production
rate of Lyman continuum photons $Q_\mathrm{Lyc}$ quantifying the SFR.
The resulting empirical calibrations provide useful tools in MIR studies
of star-forming galaxies as well as constraints for models
predicting the dust emission of such systems.

The paper is organized as follows.  Section~\ref{Sect-sample} presents the
galaxy sample.  Section~\ref{Sect-diagn} describes the MIR indicators and
the SFR estimates obtained from more classical
diagnostics.  Section~\ref{Sect-res} discusses the derived calibrations
and Sect.~\ref{Sect-conclu} summarizes the results.

\section{Galaxy sample}  \label{Sect-sample}

We drew our sample from separate studies published by us and from the ISO archive.
All sources were observed with ISOCAM either with the broadband filters 
LW2 centered at 7\,$\mu$m ($5 - 8.5\,\mu$m) and LW3
centered at 15\,$\mu$m ($12 - 18\,\mu$m) or with the continuously
variable filter (CVF) covering the $5 - 17\,\mu$m range at a
resolution of $R \equiv \lambda/\Delta\lambda \sim 40$.
Ten galaxies were observed in both broadband photometric mode and
spectrophotometric mode.  We used this subset to assess the photometric consistency
and to derive conversion factors between measurements obtained through the
ISOCAM filters and from the CVF spectra (Sect.~\ref{Sub-diagn_PAH_VSG}).
Details of the observations are given in the
relevant references (Table~\ref{tab-sample_stbs}).
To our knowledge, the ISOCAM data of IC\,342 have not been
published anywhere else; they are briefly presented in Appendix~\ref{App-IC342}.

\begin{table*}[!ht]
\caption[]{Sample of galaxies and ISOCAM observations:
           circumnuclear regions of spiral galaxies and starbursts
           \label{tab-sample_stbs}}
\setlength{\tabcolsep}{0.15cm}
\begin{center}
\begin{tabular}{lcclllll}
\hline\hline
Source & Distance\,$^{a}$ & Morph. type\,$^{b}$ & Nuclear type\,$^{c}$ &
\multicolumn{1}{c}{$L_\mathrm{IR}$\,$^{d}$} & Observations\,$^{e}$ & 
Project\,$^{f}$ & Reference$^{g}$ \\
       & (Mpc) &   &   & 
\multicolumn{1}{c}{($\mathrm{L_{\odot}}$)} &   &   &   \\
\hline
\multicolumn{8}{l}{{\em Circumnuclear regions of spiral galaxies}} \\
\object{NGC\,986} & 23.2 & SBab & \ion{H}{ii} &
 $4.6 \times 10^{10}$ & LW & Sf\_glx & 1, 3 \\
\object{NGC\,1097} & 14.5 & SBb & Liner/Sy &
 $3.8 \times 10^{10}$ & LW $+$ CVF & CAMbarre & 2 \\
\object{NGC\,1365} & 16.9 & SBb & Sy &
 $8.7 \times 10^{10}$ & LW $+$ CVF & CAMspir & 2 \\
\object{NGC\,4102} & 17.0 & SABb & \ion{H}{ii} &
 $4.2 \times 10^{10}$ & LW & Sf\_glx & 1, 3 \\
\object{NGC\,4293} & 17.0 & SB0/a & Liner &
 $4.7 \times 10^{9}$ & LW & Virgo & 4, 2 \\
\object{NGC\,4691} & 22.5 & SB0/a & \ion{H}{ii} &
 $2.4 \times 10^{10}$ & LW & CAMbarre & 2 \\
\object{NGC\,5194} (\object{M\,51}) & 7.7 & SAbc, int. & Sy &
 $2.4 \times 10^{10}$ & LW $+$ CVF & CAMspir & 5, 2 \\
\object{NGC\,5236} (\object{M\,83}) & 4.7 & SABc & \ion{H}{ii} &
 $1.9 \times 10^{10}$ & LW $+$ CVF & CAMspir & 2 \\
\object{NGC\,6946} & 5.5 & SABcd & \ion{H}{ii} &
 $1.4 \times 10^{10}$ & LW $+$ CVF & Sf\_glx, Zzcvfcam & 6, 2 \\
\object{NGC\,7552} & 19.5 & SBab & \ion{H}{ii} &
 $8.6 \times 10^{10}$ & LW & CAMbarre & 2 \\
\object{NGC\,7771}$\dagger$ & 57.2 & SBa, int. & \ion{H}{ii} &
 $2.1 \times 10^{11}$ & LW & Sf\_glx & 1, 3 \\
\hline
\multicolumn{8}{l}{{\em Starburst galaxies}} \\
\object{NGC\,253} & 2.5 & SABc & \ion{H}{ii} &
 $1.8 \times 10^{10}$ & CVF & CAMACTIV & 7 \\
\object{NGC\,520} & 27.8 & Pec, merger & \ion{H}{ii} &
 $6.5 \times 10^{10}$ & LW $+$ CVF & CAMACTIV & 8 \\
\object{NGC\,1808} & 10.8 & RSABa, int. & \ion{H}{ii} &
 $3.8 \times 10^{10}$ & LW $+$ CVF & CAMACTIV & 7 \\
\object{NGC\,3034} (\object{M\,82}) & 3.3 & I0, int. & \ion{H}{ii} &
 $4.8 \times 10^{10}$ & CVF & CAMACTIV & 7 \\
\object{IC\,342} &  3.3 & SAB(rs)cd & \ion{H}{ii} &
 $3.3 \times 10^{9}$ & CVF & IMSP\_SBG & ~ \\
\hline
\multicolumn{8}{l}{{\em LIRGs/ULIRGs}} \\
\object{NGC\,3256} & 37.4 & Pec, merger & \ion{H}{ii} &
 $4.0 \times 10^{11}$ & CVF & CAMACTIV & 8, 10 \\
\object{NGC\,6240} & 97.2 & I0:pec, merger & Liner &
 $6.0 \times 10^{11}$ & LW $+$ CVF & CAMACTIV & 8, 10 \\
\object{IRAS\,23128-5919} & 180 & merger & \ion{H}{ii} &
 $9.4 \times 10^{11}$ & LW $+$ CVF & CAMACTIV & 9 \\
\object{Arp\,220} & 72.5 & S? (Pec), merger & Liner &
 $1.3 \times 10^{12}$ & LW $+$ CVF & CAMACTIV & 8, 10 \\
\hline
\end{tabular}
\begin{list}{}{}
\item[$^{a}$] Distances are from the NGC catalogue \citep{Tul88} with
  the following exceptions:
  NGC\,253: \citet{Dav90}; 
  M\,82: \citet{Fre88}; 
  IC\,342: \citet{Sah02};
  NGC\,7771, NGC\,6240, and Arp\,220:
  computed from the \ion{H}{i} redshifts given in the RC3 catalog
  \citep[$z = 0.014300$, 0.024307, and 0.018126, respectively;][]{deV91};
  IRAS\,23128-5919: luminosity distance computed by \citet{Cha02}.
  All distances assume $H_{0} = 75~\mathrm{km\,s^{-1}\,Mpc^{-1}}$ and
  $q_{0} = 0.5$ where relevant.
\item[$^{b}$] Morphological types are from the RC3 catalogue \citep{deV91},
  with additional indications for interacting (``int.'') or merging system.
\item[$^{c}$] Nuclear types give an indication of the nuclear activity inferred
  from optical spectra.  The types for the normal spiral galaxies are as listed
  by \citet[][; see their Table~1 for references]{Rou01b}.  For a subset of
  the starburst systems, references are
  \citet[][; NGC\,253, N1808, IRAS\,23128-5919]{Kewley01},
  \citet[][; NGC\,520]{Sta91},
  \citet[][; NGC\,6240]{Vei95},
  \citet[][; Arp\,220]{Kim98}.
  The types for the remaining starbursts are those listed in the NED database.
\item[$^{d}$] Global infrared ($\mathrm{8 - 1000\,\mu m}$) luminosity 
  computed from the {\em IRAS\/} fluxes following \citet{San96}.
\item[$^{e}$] Data available from ISOCAM observations and used in our analysis:
  ``LW'' for data obtained in the broad band filters LW2 and LW3
  ($5.0 - 8.5~\mathrm{\mu m}$ and $12 - 18~\mathrm{\mu m}$),
  ``CVF'' for $5 - 17~\mathrm{\mu m}$ spectrophotometric imaging at
  $R \sim 40$.
\item[$^{f}$] Observing program.
  Sf\_glx: P.I. G. Helou; CAMbarre: P.I. C. Bonoli;
  CAMspir: P.I. L. Vigroux; Virgo: P.I. J. Lequeux;
  Zzcvfcam: P.I. D. Cesarsky;
  CAMACTIV: P.I. Mirabel; IMSP\_SBG: P.I. R. Maiolino.
\item[$^{g}$] References for initial publication of the ISOCAM data:
(1) \citet{Dal00}; (2) \citet{Rou01a}; (3) \citet{Rou01b};
(4) \citet{Bos98}; (5) \citet{Sau96}; (6) \citet{Mal96};
(7) \citet{FS03}; (8) \citet{Lau00}; (9) \citet{Cha02}; (10) \citet{Tran01}
\item[$^{\dagger}$] This galaxy also is a LIRG, but about half its total
mid-infrared emission arises outside the central regions selected here.
\end{list}
\end{center}
\normalsize
\end{table*}

The sample can be divided in four parts, in order of increasing
star formation activity:
\begin{itemize}
\item{disks of spiral galaxies,
      which form stars in a quiescent fashion and for which the relationship
      between the MIR emission and the SFR, as derived from H$\alpha$ line
      measurements, was discussed by \citet{Rou01c};}
\item{the more active circumnuclear regions of a subsample of the same
      galaxies, for which it was possible to estimate an extinction-free
      SFR from a combination of near-infrared H recombination lines
      and H$\alpha$;}
\item{nearby starburst galaxies, three of which were studied in detail
      by \citet{FS01, FS03} in their dust and fine-structure line emission;}
\item{luminous and ultraluminous infrared galaxies (LIRGs and ULIRGs,
      with $10^{11} < L_\mathrm{IR} < 10^{12}~\mathrm{L_{\odot}}$
      and $L_\mathrm{IR} \geq 10^{12}~\mathrm{L_{\odot}}$,
      respectively\footnote{$L_\mathrm{IR} \equiv L_\mathrm{8 - 1000\,\mu m}$
      is the total infrared luminosity computed from {\em IRAS\/} fluxes
      following the prescription of \citet{San96}.})
      taken from the samples studied by \citet{Lau00}, \citet{Tran01} and
      \citet{Cha02}.}
\end{itemize}

Most objects are purely star-forming systems; a few harbour an active galactic
nucleus (AGN) revealed by optical spectroscopy,
which is however known or suspected to contribute but negligibly
to the MIR emission, as well as to the hydrogen recombination lines
within our photometric apertures, which are very large with respect
to the angular size of the nucleus. Detailed notes and discussion
of individual cases are given in Appendix~\ref{App-notes}.
The LIRGs and
ULIRGs will also be referred to as starbursts throughout the paper.
Table~\ref{tab-sample_stbs} lists the non-disk objects (circumnuclear
regions of spirals and starburst systems).  The
table gives some general properties, the observation mode, and the original
observing program to which they belong.  Similar details for the spiral disk
sample are given by \citet{Rou01a, Rou01b}.

In spiral galaxies, the low brightness disks
and central regions are distinguished
by different ratios of flux density in the LW3 and LW2 filters, or
$f_\mathrm{12-18\,\mu m}/f_\mathrm{5-8.5\,\mu m}$ color: the disks
typically have ratios of $\approx 1$ while the circumnuclear regions
usually exhibit a color excess signaling more active star
formation\footnote{More generally, such a color excess can also be due
to an AGN heating a surrounding nuclear dust torus \citep[e.g.][]{Lau00}.}
\citep[e.g.][]{Dal00, Rou01b}.
We adopted the measurements for disks reported by \citet{Rou01c}.
Briefly, these were obtained from the integrated MIR and H$\alpha$ fluxes
by subtracting the contribution from a core region
and accounting for flux dilution effects of the ISOCAM point spread function
\citep[PSF; see][]{Rou01a}.  The size of the excluded area was
dictated by the H$\alpha$ data existing in the literature (fluxes in given
apertures, or maps). In the few cases where the H$\alpha$ aperture is smaller
than $D_\mathrm{CNR}$, the size of the circumnuclear regions fitted
on MIR brightness profiles, it was ensured that the resulting disk
$f_\mathrm{12-18\,\mu m}/f_\mathrm{5-8.5\,\mu m}$ color was close to unity.

Altogether, the sample covers more than five orders of magnitude in Lyman
continuum photon flux density $\Sigma_\mathrm{Lyc}$ (Sect.~\ref{Sect-res}).
The latter implies five orders of magnitude in quasi-instantaneous SFR surface
density, assuming that the same
stellar initial mass function applies to all objects \citep[e.g.][]{Ken98}.
Our sample is admittedly not complete in any sense and is restricted to
near-solar metallicities.  For our purposes, it should however provide
a sufficiently representative ensemble since the primary samples were
constructed with different criteria and aims.
Since star-forming systems, both Galactic and extragalactic,
have remarkably similar MIR spectral energy distributions (SED)
in terms of broad features and continua, we are confident that
we are not introducing any bias by selecting particular galaxies.
The sample was only shaped by the availability of adequate data.
Further details including notes on individual sources are given
in Appendix~\ref{App-notes}.

\section{Star formation diagnostics}   \label{Sect-diagn}

Tables~\ref{tab-data_MIR} and \ref{tab-data_SFR} report the data for
circumnuclear regions of spirals and starburst systems that we used
in our analysis.
The data for spiral disks are described by \citet{Rou01c}.
We reduced and analysed the MIR maps and spectra
of the whole sample in a homogeneous way.

\begin{table*}[!ht]
\caption[]{Dust emission of
           circumnuclear regions of spiral galaxies and starbursts
           \label{tab-data_MIR}}
\setlength{\tabcolsep}{0.15cm}
\begin{center}
\begin{tabular}{lccrrrrrr}
\hline\hline
Source & Region\,$^{a}$ & Area\,$^{a}$ &
\multicolumn{2}{c}{$f_\mathrm{5-8.5\,\mu m}$\,$^{b}$} & 
\multicolumn{2}{c}{$f_\mathrm{12-18\,\mu m}$\,$^{b}$} & 
\multicolumn{2}{c}{$f_\mathrm{15\,\mu m,ct}$\,$^{c}$} \\
    &   &  & LW2 & CVF & LW3 & CVF & LW & CVF \\
    & (arcsec) & ($\mathrm{pc^{2}}$) & (mJy) & (mJy) & (mJy) & (mJy) & (mJy) & (mJy) \\
\hline
\multicolumn{9}{l}{{\em Circumnuclear regions of spiral galaxies}} \\
\object{NGC\,986} & 23.6, 18.2 & $3.27 \times 10^{6}$ & 324 & ... &  673 & ... &  628 & ... \\
\object{NGC\,1097} & 45 & $7.86 \times 10^{6}$ & 1280 & (1439) & 1692 & (2200) & ... & 1641 \\
\object{NGC\,1365} & 40.0, 37.1 & $7.26 \times 10^{6}$ & 1972 & (1887) & 3103 & (3451) & ... & 2940 \\
\object{NGC\,4102} & 32.4, 28.7 & $4.39 \times 10^{6}$ & $> 571$ & ... & $\geq 1628$ & ... & 1673 & ... \\
\object{NGC\,4293} & 32.4, 6.0 & $4.39 \times 10^{6}$ & 76 & ... & 147 & ... &  134 & ... \\
\object{NGC\,4691} & 56.5, 20.4 & $2.77 \times 10^{7}$ & 542 & ... & 759 & ... &  582 & ... \\
\object{NGC\,5194} & 90 & $8.87 \times 10^{6}$ & 1883 & (2096) & 2021 & (2292) & ... & 1382 \\
\object{NGC\,5236} & 41.2, 38.4 & $6.01 \times 10^{5}$ & ($> 2753$) & 3109 & ($> 3589$) & 5468 & ... & 4369 \\
\object{NGC\,6946} & 35.3, 32.0 & $5.71 \times 10^{5}$ & ($> 1095$) & 1503 & ($> 1885$) & 2174 & ... & 1551 \\
\object{NGC\,7552} & 21.8, 19.9 & $2.79 \times 10^{6}$ & 1248 & ... & $\geq 2316$ & ... & 2067 & ... \\
\object{NGC\,7771} & 22.4, 20.6 & $2.56 \times 10^{7}$ & 314 & ... & 387 & ... &  270 & ... \\
\hline
\multicolumn{9}{l}{{\em Starburst galaxies}} \\
\object{NGC\,253} & 15.9, 14.0 & $2.27 \times 10^{4}$ & ... & 6588 & ... & 22\,755 & ... & 23\,913 \\
\object{NGC\,520} & 17.7, 15.3 & $3.36 \times 10^{6}$ & 578 & (469) & 737 & (670) & ... & 560 \\
\object{NGC\,1808} & 25.9, 24.4 & $1.28 \times 10^{6}$ & 2645 & (2450) & 4077 & (4448) & ... & 3426 \\
\object{NGC\,3034} & 30.0, 28.6 & $1.65 \times 10^{5}$ & ... & 25\,812 & ... & 62\,366 & ... & 60\,000 \\ 
\object{IC\,342}   & $17 \times 17$ & $7.40 \times 10^{4}$ & ... & 1404 & ... & 3527 & ... & 3035 \\ 
\hline
\multicolumn{9}{l}{{\em LIRGs/ULIRGs}} \\
\object{NGC\,3256} & 20.0 & $1.03 \times 10^{7}$ & ... & 1543 & ... & 2960 & ... & 2634 \\
\object{NGC\,6240} & total (3.) & $1.57 \times 10^{6}$ & 190 & (235) & 767 & (784) & ... & 801 \\
\object{IRAS\,23128-5919} & total (4.2) & $1.08 \times 10^{7}$ & 120 & (119) & 331 & (468) & 337 & (443) \\
\object{Arp\,220} & total (2.) & $2.11 \times 10^{6}$ & 191 & (196) & 765 & (869) & ... & 1170 \\
\hline
\end{tabular}
\begin{list}{}{}
\item[$^{a}$]
  ``Region'' refers to the rectangular dimensions or diameter of the
  photometric aperture; the first value is the aperture, and
  the second value is the deconvolved size, which was used for
  the area normalization as given in ``Area''.
\item[$^{b}$] Broad-band flux densities from ISOCAM observations measured
  either through the LW2 and LW3 filters or computed from
  CVF spectra accounting for the filter transmission profiles.
  Lower limits for LW measurements indicate that the nucleus is saturated
  in the maps; in which case measurements from the CVF spectrum were preferred.
  Values unused in the present analysis are enclosed in parentheses.
  From unpublished previous analysis, photometric errors are always dominated
  by incompletely corrected memory effects in the LW2 filter, and in the LW3
  filter for sources brighter than $\approx 200-500$\,mJy.
  Taking these memory effects into account, it is estimated that average
  errors are 10\% and 20\% in LW2 and LW3, respectively; individual errors
  for relatively bright galaxies may be as high as 20\% and 30\%, respectively
  \citep{Rou01a}. In addition, flux calibration uncertainties are of the order
  of 5\% (ISOCAM handbook).
\item[$^{c}$] Monochromatic 15\,$\mu$m continuum flux density derived from
  LW2 and LW3 data based on Fig.~\ref{fig-f15ct_calib} as explained
  in Sect.~\ref{Sub-diagn_PAH_VSG} or measured directly from CVF spectra.
  Calibration errors for CVFs are discussed by \citet{Biv98a}.
\end{list}
\end{center}
\normalsize
\end{table*}

\begin{table*}[!ht]
\caption[]{Line fluxes and extinction for circumnuclear regions
           of spiral galaxies and starbursts
           \label{tab-data_SFR}}
\setlength{\tabcolsep}{0.15cm}
\begin{center}
\begin{tabular}{lcccccc}
\hline\hline
Source & Aperture\,$^{a}$ & 
$F_{\lambda}$\,$^{b}$ & $F_\mathrm{H\alpha}$\,$^{b}$ & 
$A_{V}$\,$^{c}$ & $\log[Q_\mathrm{Lyc}]$~$^{d}$ &
References\,$^{e}$ \\
   & (arcsec) &
($10^{-17}~\mathrm{W\,m^{-2}}$) & ($10^{-16}~\mathrm{W\,m^{-2}}$) &
(mag) & ($\log[\mathrm{s^{-1}}]$) &  \\
\hline
\multicolumn{7}{l}{{\em Circumnuclear regions of spiral galaxies}} \\
\object{NGC\,986}   & 23.6, $4 \times 4$ & ... &  10.2 & 2.5 & 53.57 & 1, 2 \\
\object{NGC\,1097}  & 45 & ... & 39.2  & 1.9 & 53.55 & 3, 4 \\
\object{NGC\,1365}  & 40.0, 23.5 & Br$\gamma$ : 19.6  & 35.8, 28.4 & 2.9 & 53.95 & 5, 6 \\
\object{NGC\,4102}  & starburst & Pa$\beta$/Br$\gamma$ : 12.1/6.0 & ... & 7.2 & 53.55 & 7 \\
\object{NGC\,4293}  & 32.4 & Pa$\alpha$ : 10.7  &  0.76 & 3.8 & 52.59 & 8, 9 \\
\object{NGC\,4691}  & 56.5, 23.5 & Br$\gamma$ :  3.76 &  8.59, 7.18 & 2.4 & 53.45 & 10, 11 \\
\object{NGC\,5194}  & 90 & ... & 38.6  & 2.8 & 53.29 & 12, 13 \\
\object{NGC\,5236}  & 41.2, 23.5 & Br$\gamma$ : 19.6 & 75.7, 60.6  & 1.7 & 52.79 & 5, 14 \\
\object{NGC\,6946}  & 35.3, 23.5 & Br$\gamma$ : 6.1 & 8.64, 5.80 & 3.5 & 52.58 & 5, 15 \\
\object{NGC\,7552}  & 21.8, $14 \times 20$ & Br$\alpha$ : 44.0 & 30.3, 29.4 & 2.1 & 53.75 & 1, 16 \\
\object{NGC\,7771}  & starburst ring & Pa$\beta$/Br$\gamma$ : 4.0/1.5 & ... & 5.2 & 53.91 & 17 \\
\hline
\multicolumn{7}{l}{{\em Starburst galaxies}} \\
\object{NGC\,253}  & 15, $2.4 \times 12$ & Br$\gamma$ : 91.6 & ... & 8.5 & 53.13 & 18 \\
\object{NGC\,520}  & 17.7, $6 \times 8$ & Br$\gamma$ : 1.9 & 0.52, 0.15 & 7.4 & 54.03 & 19, 20 \\
\object{NGC\,1808}  & 20 (total) & Br$\gamma$ : 31.0 & ... & 4 & 53.72 & 21 \\
\object{NGC\,3034} & 30.0 & ... & ... & 52 (MIX)$^{\dagger}$ & 54.09 & 22 \\
\object{IC\,342}  & $17 \times 17$, $14 \times 20$ & Br$\gamma$ : 17.0; Br$\alpha$ : 70.0 & ... & 5.2$^{\dagger}$ & 52.48 & 16, 23 \\
\hline
\multicolumn{7}{l}{{\em LIRGs/ULIRGs}} \\
\object{NGC\,3256}  & $\approx 20$ (total), $3.5\arcsec \times 3.5\arcsec$ & Br$\gamma$ : 15.0 & ... & 5.3 & 54.54 & 24, 25 \\
\object{NGC\,6240}  & total & Br$\gamma$ : 3.1; Pa$\beta$ : 4.1 & ... & 10.1 & 54.91 & 26, 27 \\
\object{IRAS\,23128-5919}  & total, S. nucleus & Pa$\alpha$ : 8.0 & 1.76, 0.93 & 3.0 & 54.74 & 28, 29 \\
\object{Arp\,220}  & total & Br$\alpha$ : 21.0; Br$\gamma$ : 0.59 & ... & 40.1$^{\dagger}$ & 55.33 & 30, 31 \\
\hline
\end{tabular}
\begin{list}{}{}
\item[$^{a}$] Rectangular dimensions or diameter of the photometric aperture.
  When two values are given, the first one refers
  to the primary flux (H$\alpha$ when present) and the second one to the
  hydrogen line decrement used to estimate the extinction.  The aperture adopted
  for the size normalization is given in Table~\ref{tab-data_MIR}.
\item[$^{b}$] Observed line fluxes used to derive the extinction and/or
  the ionizing photon flux.
  When two H$\alpha$ fluxes are given, they correspond to the two apertures
  used for photometry and extinction estimation, respectively.
\item[$^{c}$] Derived or adopted extinction for a uniform foreground
  screen (UFS) model except when ``MIX'' indicates that a homogeneous mixture
  of dust and sources is assumed.
\item[$^{d}$] Derived or adopted intrinsic Lyman continuum photon rate.
\item[$^{e}$] References for line fluxes and extinction values
  (further details are given on the derivation of the data in
  appendix~\ref{App-notes}):
  (1) H$\alpha$+[\ion{N}{ii}] map from \citet{Ham99} (in NED); 
  (2) Extinction from H$\beta$/H$\alpha$ of \citet{Ver86} in $4\arcsec \times 4\arcsec$;
  (3) H$\alpha$+[\ion{N}{ii}] map graciously provided by T. Storchi-Bergmann \citep{Sto96};
  (4) Average extinction along the starburst ring derived from the data of \citet{Kot00};
  (5) Br$\gamma$ flux from \citet{Pux88} in an effective aperture of 23.5\arcsec;
  (6) H$\alpha$+[\ion{N}{ii}] map graciously provided by M. Naslund \citep{Kri97, Lin99};
  (7) Br$\gamma$ and Pa$\beta$ fluxes from \citet{Rou03}
      in $\approx 5\arcsec \times 5\arcsec$ (most of starburst enclosed);
  (8) Pa$\alpha$ map from \citet{Bok99} (in NED);
  (9) H$\alpha$+[\ion{N}{ii}] map from \citet{Koo01} (in NED);
  (10) Br$\gamma$ flux from \citet{Pux90} in an effective aperture of 23.5\arcsec;
  (11) H$\alpha$+[\ion{N}{ii}] map graciously provided by A. Garc\'{\i}a-Barreto \citep{Gar95};
  (12) H$\alpha$+[\ion{N}{ii}] map from \citet{Gre98};
  (13) Average extinction towards \ion{H}{ii} regions from \citet{Sco01};
  (14) H$\alpha$ map graciously provided by S. Ryder via A. Vogler \citep{Ryd95};
  (15) H$\alpha$ map from \citet{Lar99} (in NED);
  (16) Br$\alpha$ flux from \citet{Ver03} in $14\arcsec \times 20\arcsec$;
  (17) Br$\gamma$ and Pa$\beta$ fluxes from \citet{Dale03}
       in $\approx 5\arcsec \times 5\arcsec$ (most of starburst ring enclosed);
  (18) Br$\gamma$ flux and extinction from \citet{Eng98};
  (19) Br$\gamma$ flux from \citet{Sta91} in $6\arcsec \times 8\arcsec$;
  (20) H$\alpha$+[\ion{N}{ii}] map from \citet{Hibbard96} (in NED);
  (21) Br$\gamma$ flux and extinction from \citet{Kra94};
  (22) Extinction and $Q_\mathrm{Lyc}$ from \citet{FS01};
  (23) Br$\gamma$ map provided by T. B\"oker \citep{Bok97};
  (24) Total Br$\gamma$ flux from \citet{Moo94};
  (25) Extinction from Br$\gamma$/Pa$\beta$ of \citet{Doy94}
       in $3.5\arcsec \times 3.5\arcsec$;
  (26) Br$\gamma$ flux from \citet{Rieke85} in 8.7\arcsec;
  (27) Pa$\beta$ flux from \citet{Sim96} in a slit of width 1.5\arcsec;
  (28) Pa$\alpha$ flux of the southern nucleus from \citet{Kawara87};
  (29) H$\alpha$ fluxes of both nuclei form \citet{Duc97} in a slit of width 1.3\arcsec;
  (30) Br$\alpha$ flux from \citet{Sturm96} in $14\arcsec \times 20\arcsec$;
  (31) Br$\gamma$ flux of \citet{Gol95} in a slit of width 0.75\arcsec.
\item[$^{\dagger}$] Values are for the GC extinction law of \citet{Lut99a}
       when we made use of the Br$\alpha$ line at 4.05\,$\mu$m (see Sect.~\ref{Sub-diagn_H}).
\end{list}
\end{center}
\normalsize
\end{table*}

\subsection{Mid-infrared $5 - 8.5\,\mu m$ and $15\,\mu m$ emission}
            \label{Sub-diagn_PAH_VSG}

\begin{figure}[!t]
\centering
\resizebox{10cm}{!}{\includegraphics[bb=120 220 460 620,clip]{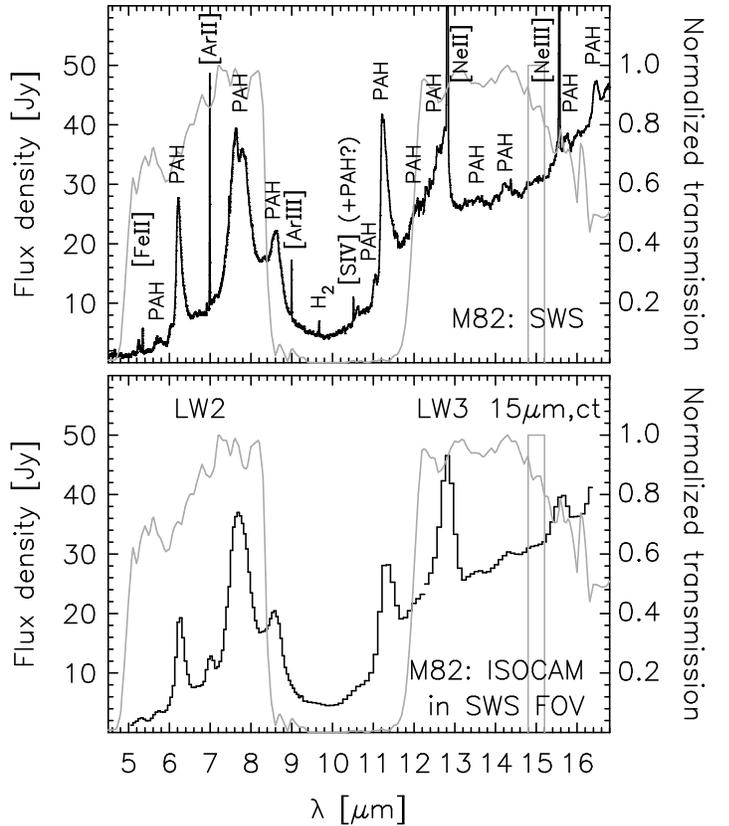}}
\caption
{
Bandpasses of the MIR indicators of star formation activity.
The transmission profiles of the ISOCAM LW2 and LW3 filters and
the narrow band used for the monochromatic 15\,$\mu$m,ct continuum
measurements are shown by the grey lines.  To illustrate the typical
spectral features covered by these bandpasses, the spectra of the
starburst galaxy M\,82 obtained at $R \sim 500 - 1000$
with the ISO SWS and at $R \sim 40$ with ISOCAM within the SWS
field of view are plotted in the top and bottom panels, respectively
(from \citealt{FS01, FS03}).  Identifications of the emission
features are labeled on the SWS spectrum.
}
\label{fig-bandpasses}
\end{figure}

\begin{figure}[!ht]
\hspace*{-1cm}
\resizebox{10cm}{!}{\rotatebox{90}{\includegraphics{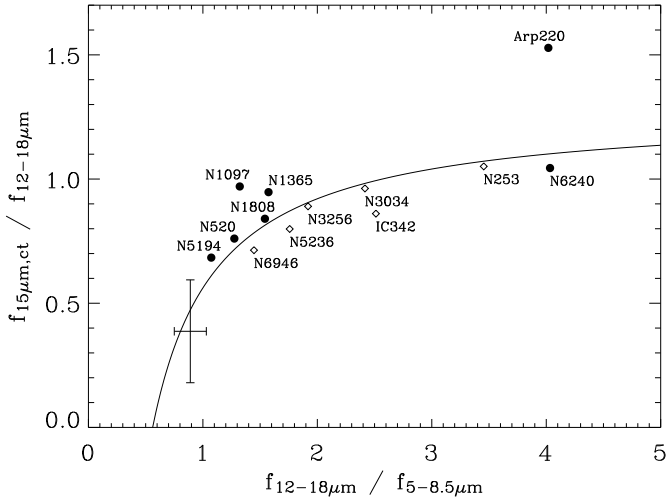}}}
\caption{
Empirical relationship used to estimate the monochromatic 15\,$\mu$m
continuum flux density from broadband LW2 and LW3 measurements.
The black circles represent circumnuclear regions of spiral galaxies
and starburst systems for which both broadband and CVF data are available.
Diamonds represent galaxies with only CVF data, for which broadband
fluxes were synthesized from the spectrum.
The line shows the least-squares fit obtained as explained in the text,
$y = a - b / x$, where $x$ and $y$ are the abscissa and the ordinate.
The error bar indicates the mean and 1$\sigma$ dispersion in
$f_\mathrm{12 - 18\,\mu m}/f_\mathrm{5 - 8.5\,\mu m}$
color of the sample of spiral disks, and the
$f_\mathrm{15\,\mu m,ct}/f_\mathrm{12 - 18\,\mu m}$ ratio and 
measurement uncertainty for the disk of NGC\,5236.
The disk of this galaxy was chosen because its spectrum has the
highest signal-to-noise ratio of the sample.
}
\label{fig-f15ct_calib}
\end{figure}

The shape of the $5 - 11\,\mu$m SED is
observed to be nearly invariant in star-forming galaxies and in a variety
of Galactic sources while at $\lambda \ga 11\,\mu$m, the substantial drop
in the most quiescent galaxies, with contributions by
minor aromatic features, contrasts with the increasingly strong and steep
continuum of VSGs in more active sources
(see Tielens 1999 for a review; see also e.g. Boulanger et al. 1998;
\citealt{Hel00, Uch00, Lau00, Stu00, Rou01c, FS03}).
In ULIRGS, extinction effects can be large enough to
distort the shape of the PAH complexes especially by the suppression of
the 8.6\,$\mu$m feature \citep[e.g.][]{Rig99}.  PAH bands are generally very
weak or absent in spectra of pure \ion{H}{ii} regions and AGNs, a fact
usually attributed to the destruction of band carriers in hard and intense
radiation fields \citep[e.g.][, and references therein]{Rig99, Lau00}.

We focussed on two bandpasses sampling as independently as possible the PAH
and VSG emission. The LW2 band ($5 - 8.5\,\mu$m) encompasses the prominent
6.2, 7.7, 8.6\,$\mu$m PAH complex where the underlying continuum emission is
generally weak in environments devoid of non-stellar activity
\citep[e.g.][]{Rig99, Lau00, Lu03}. To probe the VSG emission,
we preferred to define a narrow interval measuring
the monochromatic flux at 15\,$\mu$m rather than use
broadband measurements through the LW3 filter.  The LW3 bandpass includes
the strong 12.7\,$\mu$m PAH as well as minor features at 13.55, 14.25,
and 15.7\,$\mu$m that probably dominate the SED at low star formation
levels \citep[e.g.][]{Stu00, Hony01, Rou01c}.
We defined the 15\,$\mu$m narrow band as a top-hat profile filter
with unit transmission between 14.8 and 15.2\,$\mu$m,
maximizing the VSG contribution by avoiding known PAH features and other
possible emission lines.  We note that line emission is not expected to
contribute significantly to LW2 and LW3 measurements.  For instance,
spectra of the 
ISO Short Wavelength Spectrometer \citep[SWS;][]{deG96} at
$R \sim 1000$ show that the strongest lines falling within the
LW2 and LW3 bandpasses for M\,82 and NGC\,253
are [\ion{Ar}{ii}] 6.99\,$\mu$m, [\ion{Ne}{ii}] 12.81\,$\mu$m, and
[\ion{Ne}{iii}] 15.56\,$\mu$m \citep{Stu00, FS01};
we determined that they account for only $\approx 1$\% and
3\% of the LW2 and LW3 flux densities, respectively, in both galaxies.

Throughout this paper, we refer to the narrow 15\,$\mu$m bandpass
as ``15\,$\mu$m,\,ct'' and adopt the notations ``$5-8.5\,\mu$m'' and
``$12-18\,\mu$m'' for LW2 and LW3.  The effective bandwidths
are 0.53, 16.18, and 6.75~THz, respectively.  Figure~\ref{fig-bandpasses}
shows the corresponding wavelength ranges and transmission profiles on the
SWS spectrum of M\,82 and on the lower resolution ISOCAM spectrum
extracted within the SWS field of view \citep{FS01, FS03}.

We obtained the $f_\mathrm{5-8.5\,\mu m}$ flux densities directly from LW2
observations when available, or computed them from CVF spectra accounting
for the LW2 filter transmission profile.  Values of $f_\mathrm{5-8.5\,\mu m}$
derived from LW2 and CVF data for all the sources observed in both modes agree
within 20\%; differences may be attributed in part to possible
residuals from ghosts, flat field, and straylight in the CVF data
\citep{Biv98a, Biv98b, Oku00}.
Extrapolation of the spectra between $\approx 16-17\,\mu$m and 18\,$\mu$m
is necessary to compare the $f_\mathrm{12-18\,\mu m}$ fluxes derived from LW3 and
CVF data; however, for all the galaxies observed in both modes, we find
differences of less than 20\%, except for NGC\,1097 and IRAS23128-5919,
whose CVF data overestimate $f_\mathrm{12-18\,\mu m}$ by 30\% and 41\%,
respectively.
For NGC\,1097, this is probably because our correction of the instrument's
memory effects is most uncertain at $\lambda \geq 16\,\mu$m.
The correction algorithm \citep{Coulais00} requires knowledge of the
detector's response during previous exposures. Since CVF spectral
scans were performed in order of decreasing wavelength, and the 
illumination history of the detector prior to our observation is
unknown, the spectrum is most affected at long wavelengths.
The CVF spectrum of IRAS23128-5919 has the lowest signal to noise ratio
in the sample and is therefore less reliable; we do not use it in what follows.

Six galaxies of our sample were observed
only in broadband photometric mode, so we derived an empirical conversion between
$f_\mathrm{12-18\,\mu m}$ and $f_\mathrm{15\,\mu m,\,ct}$ as follows.
We relied on the interpretation that the emission in the LW2 band is dominated
by PAHs, and that the emission in the LW3 band is produced mainly
both by PAHs (or akin particles) and by VSGs whose flux variations behave
differently (see Sect.~\ref{Sect-intro}).
Under this assumption, $f_\mathrm{15\,\mu m,\,ct}$, which is covered by the
LW3 filter, should also contain contributions from these two species,
albeit in different proportions, and we should expect it to be related to
the broadband measurements by a simple function.
Figure~\ref{fig-f15ct_calib} shows the exact values of this function taken by
the objects for which $f_\mathrm{5-8.5\,\mu m}$ and $f_\mathrm{12-18\,\mu m}$
were measured from LW2 and LW3 maps, and $f_\mathrm{15\,\mu m,\,ct}$ from CVF
data. Assuming that the emission covered by the LW3 filter is the sum of
a component scaling linearly with the emission bands covered by LW2
and of a second component scaling linearly with the emission measured by
$f_\mathrm{15\,\mu m,\,ct}$, it is easy to show that
$f_\mathrm{15\,\mu m,\,ct} / f_\mathrm{12-18\,\mu m}$ can be represented
by an affine function of $1 / (f_\mathrm{12-18\,\mu m} / f_\mathrm{5-8.5\,\mu m})$.

Excluding Arp\,220, whose CVF spectrum suggests severe extinction effects,
with high opacity from amorphous silicates at 9.7\,$\mu$m and 18\,$\mu$m
\citep{Rig99}, and NGC\,1097 whose CVF spectrum is affected by residual
memory effects, we used the least-squares fit to these data
to assign a $f_\mathrm{15\,\mu m,\,ct}/f_\mathrm{12-18\,\mu m}$ ratio to the
circumnuclear regions of spirals and starbursts without CVF data.
Figure~\ref{fig-f15ct_calib} shows also data for galaxies observed only
in CVF mode, whose broadband fluxes were simulated from the spectrum
(diamond symbols). Except for Arp\,220 and NGC\,1097, all data points
are within 15\% of the fitted relation.
For spiral disks, we applied a uniform conversion justified by their
small dispersion in $f_\mathrm{12-18\,\mu m}/f_\mathrm{5-8.5\,\mu m}$
(\citealt{Rou01c}; see also \citealt{Dal00, Dal01}).  We used the ratio
$f_\mathrm{15\,\mu m,\,ct}/f_\mathrm{12-18\,\mu m} = 0.38$ measured on
the disk of NGC\,5236 which has the best quality CVF spectrum
among the disk sample.

We did not correct MIR flux densities for extinction.
Relative to the optical $V$ band (5500~\AA), the extinction in magnitudes
is very small, with $A_\mathrm{5-8.5\,\mu m}/A_{V}$ and
$A_\mathrm{15\,\mu m}/A_{V} \leq 0.06$ \citep[e.g.][]{Dra89, Lut99a}.
Extinction effects on the relationships studied in this work will be
discussed in Sect.~\ref{Sect-res}.

\subsection{Star formation rate indicators: H recombination lines}
            \label{Sub-diagn_SFR} \label{Sub-diagn_H}

To estimate SFRs, we used
H recombination lines collected from the literature, which provide
primary diagnostics and allowed us to derive the nebular extinction
and correct for it. We converted all fluxes
to a common reference quantity, the production rate of Lyman continuum
photons $Q_\mathrm{Lyc}$.
We took care that consistent apertures were used to measure
the dust and hydrogen line fluxes.
Limitations of available data and the assumptions on
physical conditions made in deriving $Q_\mathrm{Lyc}$ inevitably lead to
appreciable uncertainties.  We emphasize however that uncertainties
for individual sources of even a factor two affect but little
our conclusions, as will be discussed in Sect.~\ref{Sect-res}.

For spiral disks, we used the total and circumnuclear
H$\alpha$ $+$ [\ion{N}{ii}]\,$\lambda\lambda 6548,6583$~\AA\ fluxes
corrected for Galactic extinction listed by \citet{Rou01c} to obtain
disk-only fluxes. We then derived intrinsic H$\alpha$ fluxes
following the precepts of \citet{Ken83} applicable to \ion{H}{ii} regions
in spiral disks, correcting for an average 25\% contribution by the
[\ion{N}{ii}] lines and an average internal extinction
$A_\mathrm{H\alpha} = 1.1$~mag \citep[see also the discussion by][]{Rou01c}.

For the circumnuclear regions of spirals with two or more H line measurements,
we derived the extinction by least-squares fits to the ratios of observed
fluxes to intrinsic line emissivities from \citet{Sto95}.  We assigned equal
weight to the ratios given the difficulty of determining the uncertainties
for the inhomogeneous collection of H line data.  We obtained the resulting
$Q_\mathrm{Lyc}$ by averaging the individual $Q_\mathrm{Lyc}$ values derived
from each dereddened line flux, taking the total H recombination coefficient
from \citet{Sto95}.
In some cases, relevant line data were not available or too uncertain,
so we adopted published values of extinction insofar as determined from H lines.
Since the extinction was sometimes derived in a region smaller
than our photometric aperture, and the assumption of uniform extinction
throughout kiloparsec scales is probably wrong, extinction
corrections may introduce a non-negligible dispersion in the relations
shown in Sect.~\ref{Sect-res}.

Whenever
circumnuclear H$\alpha$ fluxes included the satellite
[\ion{N}{ii}] lines, we applied the same correction factor of 0.75 as for
the disks, the validity of which we verified as much as possible based on
published spectroscopy from various sources. The compilations of
\citet{Ken89}, \citet{Kennicutt92} and \citet{Jansen00} show
that while there is significant overlap in
[\ion{N}{ii}]\,$\lambda 6583$~\AA/H$\alpha$ ratios between 
disk \ion{H}{ii} regions and nuclei of spiral galaxies, many ($\sim 50$\%)
can exhibit much higher ratios, which can be explained by shock heating
(e.g. by supernova remnants) or by a non-thermal Liner/Seyfert contribution.
This line ratio increase is however observed in the 
immediate vicinity of nuclei; our circumnuclear regions are generally
much larger so that this effect is not expected to be important.

We employed the extinction law of \citet{Car89} at 
$\lambda < 3\,\mu$m and of \citet{Lut99a} at $\lambda \geq 3\,\mu$m.
We adopted a uniform foreground screen model (UFS) for the geometry of the
sources and obscuring dust.  The limited number of H lines considered for
each galaxy prevented us from constraining the extinction model,
but computations for a homogeneous mixture of dust and sources (``MIX'' model)
imply $Q_\mathrm{Lyc}$ values differing by at most 56\% (on average 18\%) from
those of the UFS model\footnote{The observed and intrinsic line fluxes
$F_{\lambda}$ and $F^{0}_{\lambda}$ are related through
$F_{\lambda} = F^{0}_{\lambda}\,\mathrm{e}^{-\tau_{\lambda}}$
for the UFS model and
$F_{\lambda} = 
F^{0}_{\lambda}\,[(1 - \mathrm{e}^{-\tau_{\lambda}})/\tau_{\lambda}]$
for the MIX model, where $\tau_{\lambda}$ is the optical depth of the
obscuring material and the corresponding extinction in magnitudes is
$A_{\lambda} = 1.086\,\tau_{\lambda}$.}.
We assumed that the \ion{H}{ii} regions in all sample galaxies are ionization
bounded, optically thick in the Lyman lines and optically thin in all others
(case B recombination), and adopted electron density and temperature of
$n_\mathrm{e} = 100~\mathrm{cm^{-3}}$ and $T_\mathrm{e} = 5000$~K.
These $n_\mathrm{e}$ and $T_\mathrm{e}$ were found representative for a
sample of starbursts observed with SWS by \citet{Tho00}, including most
starbursts in our own ISOCAM sample.  Higher values up to
$n_\mathrm{e} = 10^{4}~\mathrm{cm^{-3}}$ and $T_\mathrm{e} = 10^{4}$~K
may be more appropriate for \ion{H}{ii} regions in normal spiral galaxies of
near-solar metallicity \citep[e.g.][]{Smi75, Sha83, Giv02}.  However, the
computations of \citet{Sto95} imply variations of the relative emissivities
of 13\% on average (25\% at most) for the lines considered here, little
affecting the extinction estimates ($< 0.5$~mag for the UFS model).
The mean increase in the derived $Q_\mathrm{Lyc}$ between
$n_\mathrm{e} = 100~\mathrm{cm^{-3}}$, $T_\mathrm{e} = 5000$~K and
$n_\mathrm{e} = 10^{4}~\mathrm{cm^{-3}}$, $T_\mathrm{e} = 10^{4}$~K
is 36\% (maximum 55\%), mainly driven by the variations of the total H
recombination coefficient $\alpha_\mathrm{B}$ with $T_\mathrm{e}$
\citep[$\alpha_\mathrm{B}$ depends only weakly on $n_\mathrm{e}$;][]{Sto95}.
In addition, we do not consider individual bright \ion{H}{ii} regions,
but the total emission from large areas encompassing many star formation
complexes. The average values of $n_\mathrm{e}$ and $T_\mathrm{e}$
are thus expected to be much lower than in resolved \ion{H}{ii} regions.

\subsection{Size normalization}   \label{Sub-diagn_size}

To obtain scale- and distance-independent quantities, we normalized each
measurement by the projected surface area and expressed the results
in $\mathrm{L_{\odot}~pc^{-2}}$ (denoted hereafter
$\Sigma_\mathrm{5-8.5\,\mu m}$, $\Sigma_\mathrm{15\,\mu m,\,ct}$,
$\Sigma_\mathrm{Lyc}$).  We chose these units specifically to avoid
artificial correlation due to scale effects whereby the brighter (larger)
galaxies tend to be brighter at all wavelengths.  Normalizing all three
quantities by the surface area eliminates dispersion
from uncertainties in distance estimates.  The $Q_\mathrm{Lyc}$ values
were transformed into Lyman continuum luminosities $L_\mathrm{Lyc}$ assuming
an average ionizing photon energy of 16~eV. We emphasize that although
these quantities are formally equivalent to surface brightnesses, they
are not intended as such and will be referred to as
``size-normalized luminosities.''

We normalized fluxes of spiral disks by the circular area of
diameter $D^{B}_{25}$, the major axis length
of the $B$-band isophote $\mu_{B} = 25~\mathrm{mag\,arcsec^{-2}}$
\citep[from the RC3 catalog;][]{deV91}.  This area encloses all
detected MIR emission as defined by the LW2 isophote at
$5\,\mathrm{\mu Jy\,arcsec^{-2}}$, the typical depth reached in these
ISOCAM data, with isophote diameter ratios 
$D^{5-8.5\,\mathrm{\mu m}}_{5\,\mathrm{\mu Jy}}/D^{B}_{25}$
in the range $0.35 - 1$, depending on the gas richness and inclination
of galaxies \citep{Rou01a}.
For galaxies with no available H$\alpha$ map, it was verified that the
aperture of the integrated H$\alpha$ flux is
larger than or comparable to the spatial extent of the MIR emission.
The mismatch between the optical diameter and the actual sizes of
the MIR and H$\alpha$ emitting regions will introduce either a 
``correlation bias,'' whereby points move along a line of slope 1, or scatter
in our relationships, depending on how well
the MIR and H$\alpha$ emission trace each other and are covered by the
photometric apertures.

For the circumnuclear regions and starburst cores,
we have striven to adopt as well a uniform and well-defined surface quantity.
We extracted the size of the MIR source from azimuthally-averaged
surface brightness profiles fitted with a gaussian, after bringing all images
to a common gaussian PSF (see Appendix~\ref{App-notes}). We chose a
photometric aperture of 2.5 times the fitted half-power beam width
(HPBW) in order to measure
the total flux to a good aproximation; after deconvolution by the size
of the PSF, the aperture is also used for area normalization.
For the sources whose geometric structure is ill-suited to such a
definition, we made exceptions explained in Appendix~\ref{App-notes}.
In particular, for the most compact sources (with respect to the ISOCAM
angular resolution), NGC\,4293, NGC\,6240, IRAS23128-5919
and Arp\,220, we obtain in this way intrinsic sizes that are comparable to
or smaller than the angular resolution, and therefore highly uncertain. Hence,
for these galaxies, we did not use the fitted MIR sizes for area normalization,
but starburst sizes derived from high-resolution observations in the literature.

\section{Results}   \label{Sect-res}

\subsection{Calibration of MIR dust emission as star formation diagnostic}
            \label{Sub-res_calib}

\begin{figure}[!ht]
\hspace*{-1cm}
\resizebox{10cm}{!}{\rotatebox{90}{\includegraphics{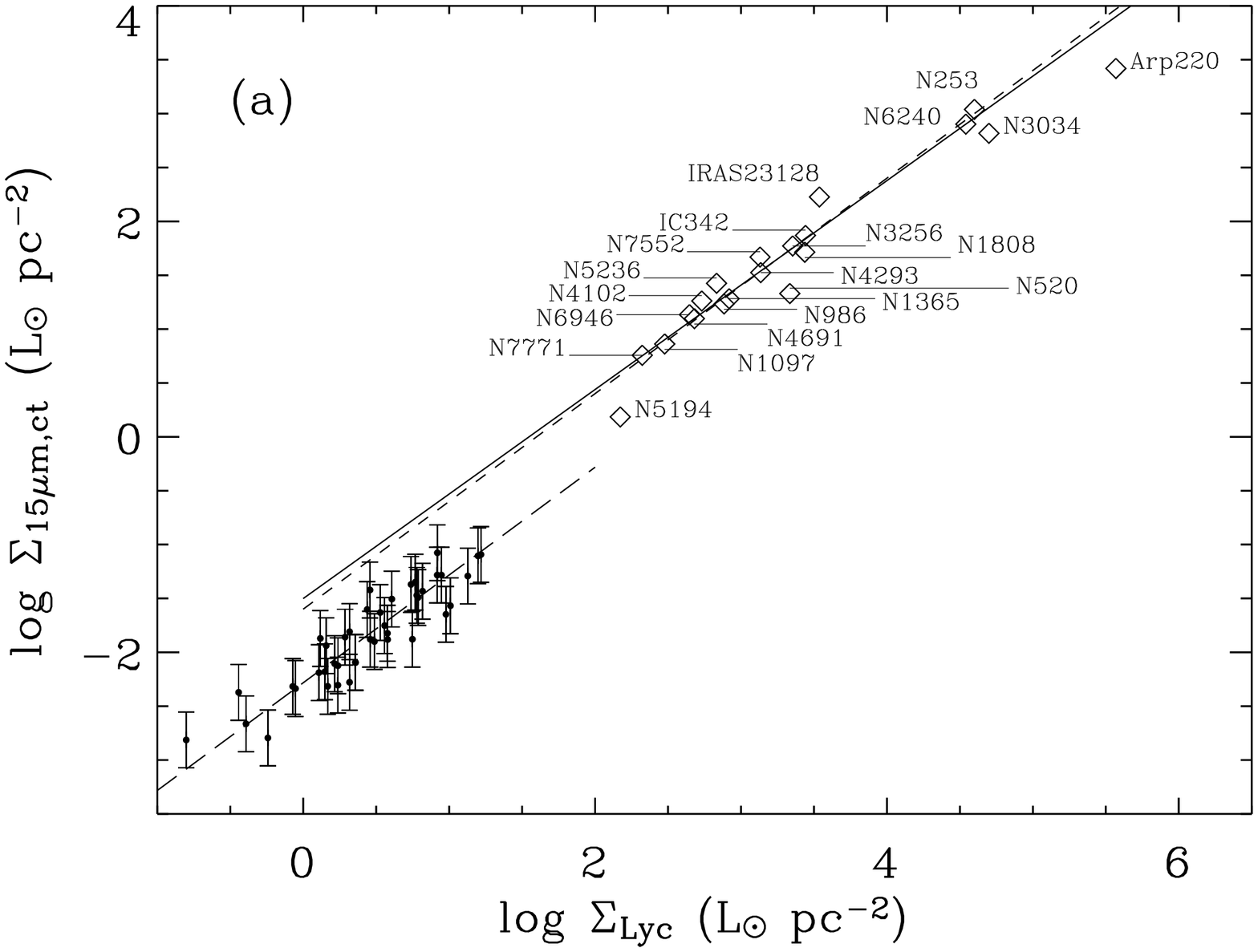}}}
\vspace*{-0.5cm} \\
\hspace*{-1cm}
\resizebox{10cm}{!}{\rotatebox{90}{\includegraphics{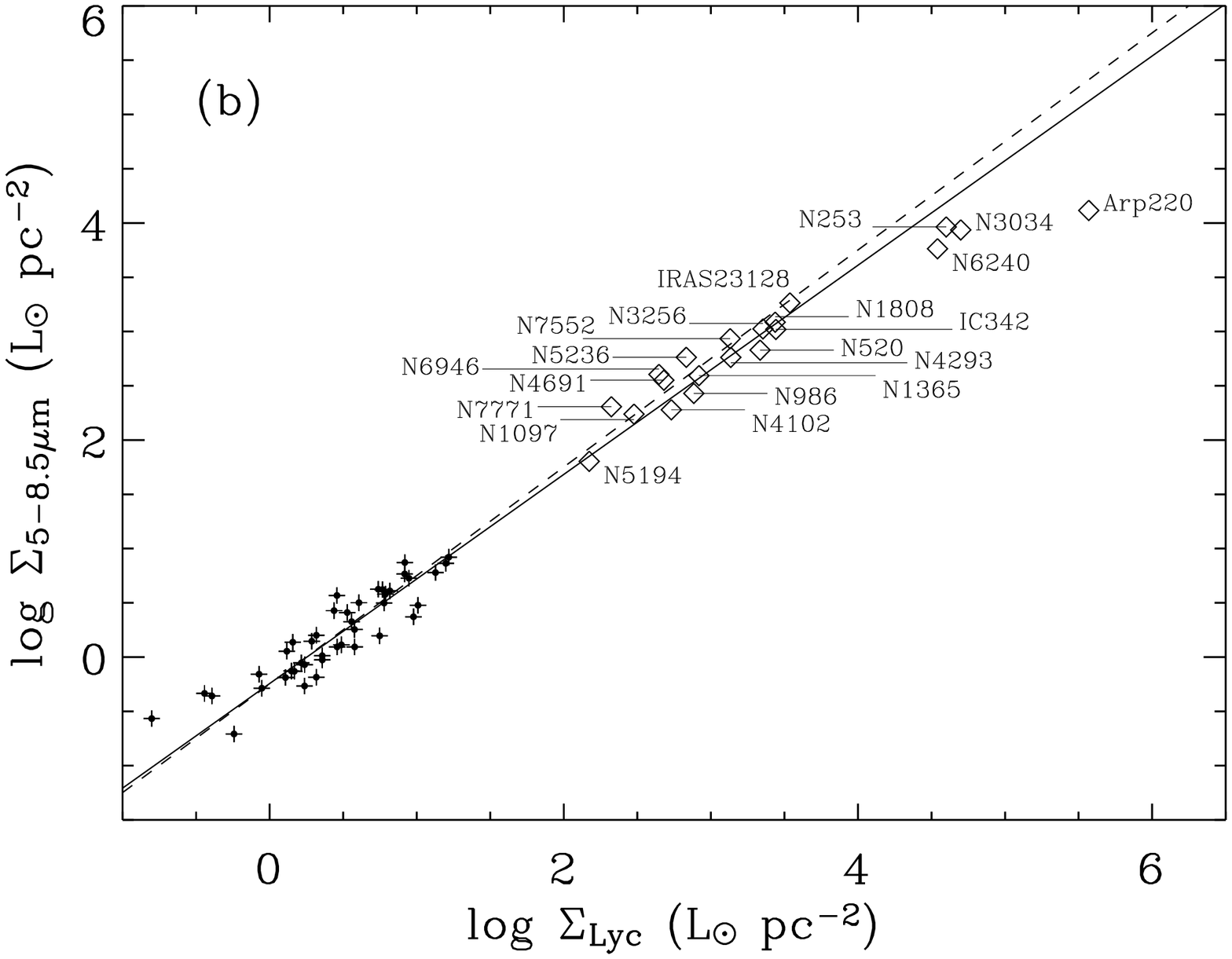}}}
\caption
{
Empirical relationships between the MIR dust emission and the
Lyman continuum luminosity as a measure of the star formation rate.
The quantities plotted are size-normalized luminosities, with different
symbols used for the spiral disks (crosses), and the circumnuclear regions
of spirals and starburst systems (diamonds).
{\bf (a)} In the $\log(\Sigma_\mathrm{15\,\mu m,ct}) - \log(\Sigma_\mathrm{Lyc})$
diagram, the solid line indicates the power-law least-squares
fit performed on the circumnuclear regions and starbursts only, excluding
the disks, NGC\,5194 and Arp\,220. 
The short-dashed line shows the fit result where the index was fixed to unity,
and the long-dashed line shows a similar fit for disks alone.
{\bf (b)} In the $\log(\Sigma_\mathrm{7\,\mu m}) - \log(\Sigma_\mathrm{Lyc})$ diagram,
the conventions are the same and fits were performed including the disks,
but excluding NGC\,253, NGC\,3034 (= M\,82), NGC\,6240 and Arp\,220.
}
\label{fig-f15_lw2_qlyc}
\end{figure}

Figure~\ref{fig-f15_lw2_qlyc} shows the relationships between the observed
$\Sigma_\mathrm{15\,\mu m,ct}$ and $\Sigma_\mathrm{7\,\mu m}$ and the derived
$\Sigma_\mathrm{Lyc}$ for our sample galaxies.  The immediate result is that
both MIR quantities constitute reliable SFR tracers
over many orders of magnitude in $\Sigma_\mathrm{Lyc}$.
Thus, in a general sense, our results extend those previously found for
spiral disks by \citet{Rou01c} and for NGC\,253, NGC\,1808,
and M\,82, including individual regions, by \citet{FS03}.
Our enlarged sample reveals however interesting differences in the
behaviour of the MIR tracers and between various source classes.

In the $\Sigma_\mathrm{15\,\mu m,ct} - \Sigma_\mathrm{Lyc}$ diagram,
the monochromatic 15\,$\mu$m continuum emission is directly
proportional to the ionizing photon luminosity within the error bar on
the power-law index, but with two different
normalizations corresponding to two distinct regimes, each spanning
$2 - 3$ orders of magnitude in $\Sigma_\mathrm{Lyc}$:
1) quiescent spiral disks, with low Lyman continuum luminosities per unit
projected area $\Sigma_\mathrm{Lyc} \la 10^{2}~\mathrm{L_{\odot}\,pc^{-2}}$;
and 2) moderately to actively star-forming regions in the central 
$\mathrm{\la 1~kpc}$ of spiral and starburst galaxies, with an activity level
characterized by $\Sigma_\mathrm{Lyc} \ga 10^{2}~\mathrm{L_{\odot}\,pc^{-2}}$,
but perhaps excluding extreme environments with
$\Sigma_\mathrm{Lyc} \sim 10^{5}-10^{6}~\mathrm{L_{\odot}\,pc^{-2}}$.
Although the transition between the two regimes is certainly gradual,
our data set includes but one object sampling it. It occurs approximately
at the level of star formation activity seen in the inner
$\approx 90$\arcsec\ plateau of M\,51, which has a color
$f_\mathrm{12 - 18\,\mu m}/f_\mathrm{5 - 8.5\,\mu m} \approx 1.1$.
The offset between the normalization of disks and that of starbursts
is nearly a factor 5, and cannot be caused by the different methods
applied to estimate ionizing photon fluxes.
Assuming direct proportionality in each separate regime, we obtain:
\begin{eqnarray}
  \log(\Sigma_\mathrm{15\,\mu m,ct}) = \log(\Sigma_\mathrm{Lyc}) - 2.28, &
  \log(\Sigma_\mathrm{Lyc}) < 2, \\
  \log(\Sigma_\mathrm{15\,\mu m,ct}) = \log(\Sigma_\mathrm{Lyc}) - 1.60, &
  2 \leq \log(\Sigma_\mathrm{Lyc}) < 5, 
\label{f15ct_qlyc_fits}
\end{eqnarray}
where all size-normalized luminosities are in $\mathrm{L_{\odot}\,pc^{-2}}$.
The fits are shown as dashed lines in Fig.~\ref{fig-f15_lw2_qlyc}.
The dispersions are, respectively,
0.22~dex for the disks (factor $\approx 1.6$), and 0.18~dex
for the galactic centers and starburst systems (factor 1.5),
from which we have excluded NGC\,5194 and Arp\,220 at the extremes
of the $\Sigma_\mathrm{Lyc}$ range.
Linear least-squares fits, where the power-law index is let as a free parameter,
yield an exponent of $1.01 \pm 0.07$ for disks and $0.97 \pm 0.06$ for starbursts.

The break between spiral disks and more active regions can be
interpreted easily.  In disks, the density of the radiation heating the dust
is too low for the VSG continuum to be significant at 15\,$\mu$m.  In such
conditions, the continuum starts rising at longer wavelengths where
larger dust grains at lower temperatures re-emit the energy absorbed from the
relatively diffuse radiation field.  The emission detected at 15\,$\mu$m is then
dominated by PAHs or a related family of particles.
Above a certain threshold in ionizing radiation density, VSG heating
becomes more efficient such that the continuum starts to make a significant
contribution at 15\,$\mu$m, and another regime prevails.  The break thus
signals the onset of VSG emission at sufficient star formation densities.
Since linear correlations represent adequately the data in both regimes,
we infer that the respective contributions from each dust species vary
almost linearly with Lyman continuum luminosity.  The data of
Arp\,220 suggest a flattening of the relationship at the most extreme
densities; its $\Sigma_\mathrm{15\,\mu m,ct}$ lies about a factor of 4
below a simple extrapolation of the linear correlation seen in the other
starbursts and the circumnuclear regions.
Since the MIR spectrum of Arp\,220 shows some signs of high optical depth
\citep{Rig99}, this damping of the dust emission can easily be caused
by extinction effects.

At 7\,$\mu$m, the situation is much different. Perhaps surprisingly, the PAH
emission remains as good a star formation indicator in circumnuclear regions
and starbursts as in spiral disks.
The linear relationship previously defined by the disks alone holds up to
$\Sigma_\mathrm{Lyc} \approx 10^{4}~\mathrm{L_{\odot}\,pc^{-2}}$;
adopting this value as transition point and assuming direct proportionality,
we obtain:
\begin{eqnarray}
  \log(\Sigma_\mathrm{7\,\mu m}) = \log(\Sigma_\mathrm{Lyc}) - 0.25, &
  0 \leq \log(\Sigma_\mathrm{Lyc}) < 4,
\label{f7um_qlyc_fits}
\end{eqnarray}
where again the size-normalized luminosities are in
$\mathrm{L_{\odot}\,pc^{-2}}$.  The dispersion of circumnuclear regions and starbursts
is 0.17~dex when galaxies beyond $\Sigma_\mathrm{Lyc} = 10^{4}~\mathrm{L_{\odot}\,pc^{-2}}$
are excluded, and increases to 0.25~dex (a factor 1.8) when the same objects
as for the $\Sigma_\mathrm{15\,\mu m,ct}$--$\Sigma_\mathrm{Lyc}$ relation
are considered. The dispersion of disks is 0.21~dex.
Allowing the power-law index to vary, the least-squares fit over the entire
$0 \leq \Sigma_\mathrm{Lyc} < 4$ range gives an exponent of $0.96 \pm 0.02$.

The 7\,$\mu$m fluxes start to deviate
significantly from the extrapolation of the linear correlation defined by
disks above $\Sigma_\mathrm{Lyc} \approx 10^{4}~\mathrm{L_{\odot}\,pc^{-2}}$.
The starburst cores of NGC\,253, M\,82 and NGC\,6240
fall by a factor 2--3 below the expected values while
Arp\,220 lies more than an order of magnitude lower.
Extinction effects alone cannot account for the saturation of the
7\,$\mu$m emission beyond $\Sigma_\mathrm{Lyc} = 10^{4}~\mathrm{L_{\odot}\,pc^{-2}}$,
which is most probably caused by disappearance of the band carriers
from the starburst cores (see Sect.~\ref{Sect-conclu}).
This is not in contradiction with the different
relation found at 15\,$\mu$m: the fact that VSGs are larger and more
resilient than PAHs allows the 15\,$\mu$m diagnostic to continue rising up
to higher star-formation activity levels than the 7\,$\mu$m emission,
though eventually VSG destruction might become significant too.

Although extinction is not expected to be a dominant cause of the 7\,$\mu$m
emission deficit at high values of $\Sigma_\mathrm{Lyc}$,
it could, however, increase the dispersion at the highest
star formation rate densities, together with variations in the average physical
conditions of the gas (electronic densities and temperatures) and in metallicity.
Another, certainly more important source of scatter is due to limitations
of the available H line measurements and data used to estimate the nebular
extinction. In particular, although we tried to minimize such effects as much
as possible, the apertures are not perfectly coincident, the angular
resolutions are not perfectly matched, and the extinction correction was
sometimes derived in a region much smaller than the aperture.
In view of all the uncertainties arising from use of inhomogeneous
data and from assumptions about physical conditions in \ion{H}{ii} regions,
and considering the fact that the observed dispersions are very
small compared to the dynamic range of our relations, we have
demonstrated that the two dust tracers investigated here constitute
satisfactory and quantitative star formation estimators.
We insist that the galaxies included in our sample are all
of near-solar metallicities, while the relation between ionizing photon
luminosity and dust emission may be very sensitive to a decrease
in carbon abundance.

\subsection{MIR color as an indicator of compactness}
            \label{Sub-res_colour}

\begin{figure}[!t]
\hspace*{-1cm}
\resizebox{10cm}{!}{\rotatebox{90}{\includegraphics{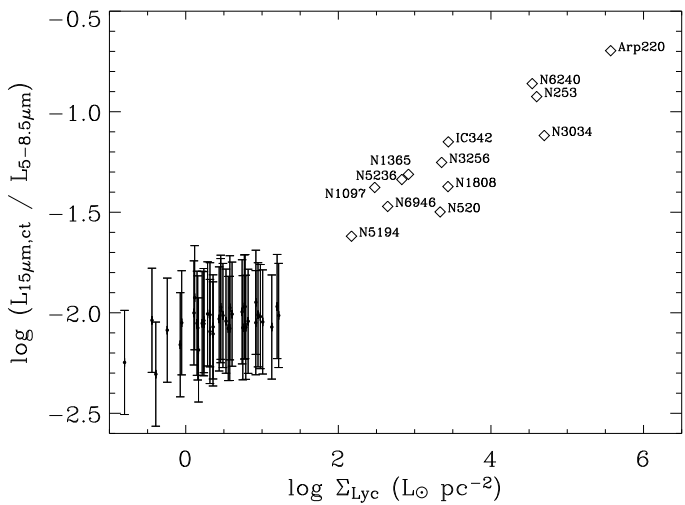}}}
\caption{
Evolution of the 
$L_{\rm 15\,\mu m,ct}/L_\mathrm{5-8.5\,\mu m}$ ratio
with increasing Lyman continuum luminosity per unit projected surface area.
Apart from the disks, only galaxies with a CVF spectrum are shown here.
}
\label{fig:nlyc_color}
\end{figure}

Since the two dust tracers behave differently (the
linearity ranges and dispersions of the relations discussed above are
different), their variations relative to each other may provide useful
diagnostics on the star formation activity. Figure~\ref{fig:nlyc_color}
shows how the $L_{\rm 15\,\mu m,ct}/L_\mathrm{5-8.5\,\mu m}$
ratio varies with increasing $\Sigma_\mathrm{Lyc}$. It should be
noted here that the uncertainty on the actual size of the emitting
regions, which affects only the abscissa, is potentially quite large.
It is however expected to be at most a factor of a few, i.e. very small
compared with the variation amplitude of $\Sigma_\mathrm{Lyc}$, which
is three orders of magnitudes for the sole starburst regions. We find
that the $\Sigma_{\rm 15\,\mu m,ct}/\Sigma_\mathrm{5-8.5\,\mu m}$ ratio,
tracing to the first order the ratio of VSG emission to PAH emission,
increases regularly from disks to mild starbursts to extreme starbursts.
A similar relationship was found by \citet{Dale03} in nuclear regions
and extranuclear \ion{H}{ii} regions of spiral galaxies. Their ionizing
photon flux densities, derived in a uniform way from integral-field
Pa$\beta$ and Br$\gamma$ lines, correspond to $\Sigma_\mathrm{Lyc}$,
the quantity used here, between $10^{2.5}$ and $10^4~\mathrm{L_{\odot}\,pc^{-2}}$.
Although with a large dispersion, in part because of the inhomogeneous
nature of the data used and the moderate angular resolution in the infrared,
we show that this trend exists over a much larger range of star formation
rate density.

$\Sigma_\mathrm{Lyc}$ is essentially a quantification of the compactness
of star formation activity, and is affected both by geometry effects
(filling factor of the interstellar medium by \ion{H}{ii} complexes)
and by excitation effects (mass spectrum of ionizing stars). These
two effects produce qualitatively similar results on the
$\Sigma_{\rm 15\,\mu m,ct}/\Sigma_\mathrm{5-8.5\,\mu m}$ ratio, the PAH
emission being reduced because the sites of emission (mainly the surface of
molecular clouds) collectively decrease in volume relative to
\ion{H}{ii} regions, and because the PAH carriers may be destroyed over large
spatial scales when the radiation field intensity becomes high enough.
Excitation and density effects are intimately connected \citep{Dale03},
but variations in radiation hardness (or age of the dominant stellar
populations) are also expected to increase the color scatter,
as was found by \citet{Rou01b} for central regions of spiral galaxies.
The influence of metallicity differences is expected to be negligible
for this particular sample, and differential extinction effects are very
small compared with the dispersion, except perhaps for Arp\,220.

\section{Summary and conclusions}    \label{Sect-conclu}

At mid-infrared wavelengths, dust species emitting aromatic bands
seen mainly in the 6-13\,$\mu$m range and a continuum
rising toward longer wavelengths provide important observables
for studies of star formation in dusty environments, and provide
finer details than the far-infrared emission of grains in thermal
equilibrium, because of much more favorable angular resolution
and lesser source confusion.
The fraction of the total infrared power produced by each of these
two species is of the order of 10--20\%, depending on the excitation
conditions of dust grains \citep{Dal01, Dale02}; the fraction contributed
by very small grains, in particular, is remarkably constant up to very
high average temperatures.
In order for the SFR calibration at 5-8.5\,$\mu$m derived here to be
consistent with the calibration in terms of total infrared emission
(between 8 and 1000\,$\mu$m) proposed by \citet{Ken98} for starbursts,
the power emitted in the 5-8.5\,$\mu$m range has to amount to 18\%
on average of the total infrared power (for galaxies less active than
M\,82). Owing to the fact that the far-infrared emission of galaxies
is not spatially resolved by IRAS, and that the central regions
selected here emit only a fraction of the integrated MIR emission
of each galaxy, we cannot rigorously estimate the part of aromatic bands
in the energy budget separately for disks and for galactic centers
(but this will become feasible in local galaxies with observations by the
Spitzer satellite). We only remark that a power fraction of 18\% is somewhat
larger than the fractions inferred by \citet{Dal01} for a wide range
of dust temperatures.

We have investigated the response of these dust species to the
radiation field generated by massive stars, estimated independently
and corrected for extinction, in a sample of star-forming sources
of near-solar metallicity. In our sample, ionizing photon flux densities
span a very wide range, from $\approx 1$ to $\approx 10^{5.5}~\mathrm{L_{\odot}\,pc^{-2}}$.
The regions considered here are spiral disks on one hand, representative
of quiescent environments, and circumnuclear regions on the other hand,
extended on spatial scales of the order of the kiloparsec.

Even though aromatic band carriers are on average heated by softer radiation
than very small grains, we have shown that they can be used as a
quantitative star formation tracer, their emission scaling linearly
with the intrinsic emission of hydrogen recombination lines over a dynamic
range of four orders of magnitude in ionizing photon flux densities.
The relation found here confirms and extends that previously found for
spiral disks up to much higher star formation rate densities. The global
emission from aromatic bands starts to be damped past activity levels
only just milder than that of M\,82. By analogy with what is observed
in and around \ion{H}{ii} regions in the Galaxy and the Magellanic
Clouds, this saturation is most probably caused by the gradual destruction
of aromatic band carriers effected by more and more intense far-ultraviolet
radiation fields \citep{Tran98, Con00}. In fact, this may be an indirect
cause, \citet{Giard94} having found a tighter relationship of
the 3.3\,$\mu$m PAH brightness with the ionized gas density than with
the radiation field intensity. Additional agents of
dust grain destruction may be found in the enhanced cosmic ray density
from numerous supernova explosions \citep{Mennella97}, and in
high-velocity starburst winds \citep{Normand95}.

Such a behavior as seen here in galaxies was previously reported for individual
photodissociation regions by \citet{Boulanger98}. The approximate threshold
at which they observe a significant depletion of aromatic bands
is $\approx 10^{3.5}$ times the radiation field of the solar neighborhood $G_0$,
or $10^{4.2}~\mathrm{L_{\odot}\,pc^{-2}}$ in the unit used here
\citep[adopting $G_0 = 2.2 \times 10^{-6}$\,W\,m$^{-2}$ from][]{Mathis83}.
The threshold applicable to galactic starburst regions, occurring around
$10^{4 \pm 0.5}~\mathrm{L_{\odot}\,pc^{-2}}$, is fully consistent with
that found by \citet{Boulanger98}.
It should be noted that resolution and dilution effects,
as well as incomplete sampling of the explored radiation field range,
hamper equally both studies, so that the actual value of the
threshold is somewhat uncertain. The collective behavior of star-forming
regions, integrated over kiloparsec scales, is nevertheless
similar to that of individual \ion{H}{ii} regions and the associated
neutral material surrounding them.
This result suggests that the volume ratio of ionized regions on one hand,
and surrounding regions where aromatic bands are excited on the other,
does not vary in a systematic way up to the above mentioned radiation
field intensity threshold, and then increases steadily, ionized regions
occupying a growing fraction of the interstellar medium and starting
to overlap.

The continuum of very small grains (sampled at 15\,$\mu$m), on the other hand,
provides a star formation rate tracer that is valid at higher radiation
field intensities. In practice, variations of the spectral energy
distribution of very small grains with their temperature distribution
may cause appreciable deviations according to the sampled wavelength range,
but we have shown here that the proportionality between ionizing
photon fluxes and the 15\,$\mu$m continuum is impressively tight,
as soon as the VSG continuum dominates the bandpass, and at
least up to $\Sigma_\mathrm{Lyc} = 10^{5}~\mathrm{L_{\odot}\,pc^{-2}}$.
Very small grains may also be destroyed in very harsh radiation fields
\citep{Con00}, but this effect is not observed here except possibly in Arp\,220,
where it is however not separable from optical depth effects.

New space missions such as Spitzer are making the mid-infrared window ever
more accessible and are going to perform large surveys of galaxies.
The choice to measure the continuum of very small grains at 15\,$\mu$m
was dictated by the limited wavelength coverage of the data we used.
However, with the Spitzer satellite, this continuum will be observable
primarily through the 20-28\,$\mu$m filter of the MIPS instrument,
and through the IRS Long-Low spectrometer for brighter galaxies.
For local galaxies, these wavelengths promise an excellent star formation
tracer, following a single regime from disk-like to very high radiation
field intensities. The MIPS 24\,$\mu$m filter will detect the continuum
of very small grains from $z = 0$ to $z \approx 0.6$, shifting gradually
down to 15\,$\mu$m, then the aromatic band cluster at 6-9\,$\mu$m at
$z \approx 1.8$--2.7\,. The MIPS 70\,$\mu$m filter will cover the continuum
of very small grains from 30\,$\mu$m to 15\,$\mu$m from $z \approx 1.3$ to
$z \approx 3.5$\,. The quantitative relationships that we have derived
in this paper might thus prove very useful in the immediate future.

\begin{acknowledgements}

Our referee, Dr. D.A. Dale, is gratefully thanked for his swiftness
and help in improving the dicussion flow.
It is a pleasure to thank all the persons who made some of the data
used here available to us or publicly (and who are named in Table~3).
V.C. would like to acknowledge the support of JPL contract 960803.
This research made use of the NASA/IPAC Extragalactic Database (NED)
which is operated by the Jet Propulsion Laboratory, California Institute
of Technology, under contract with the National Aeronautics and Space
Administration.
The ISOCAM data presented in this paper were analyzed using and adapting the CIA
package, a joint development by the ESA Astrophysics Division and the ISOCAM
Consortium (led by the PI C. Cesarsky, Direction des Sciences de la Mati\`ere,
C.E.A., France).

\end{acknowledgements}

\appendix

\section{Notes on individual sources and photometry}   \label{App-notes}

We give here additional details concerning individual sources as well as
MIR and line flux measurements, for the spiral galaxies whose
circumnuclear regions are studied in this work and the starburst systems.
Whenever the H$\alpha$ data included a contribution by the adjacent
[\ion{N}{ii}]\,$\lambda\lambda\,6548,6583\,\mathrm{\AA}$ lines,
we applied a correction factor of 0.75 to obtain pure H$\alpha$ fluxes
(see Sect.~\ref{Sub-diagn_H}).  ``[\ion{N}{ii}]'' refers to both lines
at $\lambda = 6548$ and 6583~\AA, except if the wavelength of the line
actually meant is given.
Unless explicitely specified otherwise, extinction values derived from H line
measurements are referred to a uniform foreground screen (UFS) model,
case B recombination, with $n_{\rm e} = 100~\mathrm{cm^{-3}}$, and
$T_{\rm e} = 5000~\mathrm{K}$.

We brought the MIR maps in all bandpasses
to the same angular resolution before performing aperture photometry,
substituting the extended non-gaussian PSF with a gaussian PSF of
FWHM 6\arcsec\ (when the pixel size is 6\arcsec), 3.5\arcsec\
(when the pixel size is 3\arcsec) or 3\arcsec\ (when the pixel size
is 1.5\arcsec). To do this, we used an iterative procedure with a gain
of 5\% to ensure convergence, centered on the brightest pixel at each step.
When using a map to measure hydrogen recombination line fluxes,
we then convolved this map to the same angular resolution as in the MIR.
Except when contrary indication is given below, we chose an homogeneous
definition of the aperture as 2.5 times the half-power beam width fitted
on the central MIR brightness profiles; this provides a very good
approximation of the total flux of the central regions, coinciding
well with sizes reported in \citet{Rou01b} (obtained by decomposing
brightness profiles into a gaussian core and an exponential disk).
\\

{\em NGC\,986} --
We assumed that the extinction derived from the H$\beta$/H$\alpha$
ratio in the central $4\arcsec \times 4\arcsec$ from
\citet{Ver86} is representative of that in our larger aperture.
The data of \citet{Ver86} imply 
[\ion{N}{ii}]\,$\lambda\,6583$\,\AA/(H$\alpha$ + 
[\ion{N}{ii}]\,$\lambda\,6583$\,\AA) $= 0.30$ at the nucleus.

{\em NGC\,1097} --
This strongly barred Liner/Seyfert galaxy has a bright star-forming ring of
diameter $\approx 20$\arcsec\ \citep[e.g.][]{Hum87, Kot00}.
The nucleus, which is resolved and separable from the ring in our maps,
contributes negligibly to the integrated H$\alpha$ and Br$\gamma$
emission \citep{Sto96, Kot00}, as well as to the total MIR emission.
We used an aperture of 45\arcsec, encompassing the whole emission
from the ring \citep{Rou01b}.
We corrected for the average extinction based on the results of
\citet{Kot00} derived from H$\alpha$/Br$\gamma$ ratios.
We recomputed the weighting by the Br$\gamma$ luminosity, with
adjustments for the [\ion{N}{ii}] contribution to their H$\alpha$ data
and the different extinction laws adopted.

{\em NGC\,1365} --
We combined the Br$\gamma$ measurement of \citet{Pux88} with the H$\alpha$ flux
integrated within the same region from an H$\alpha$ $+$ [\ion{N}{ii}] map
to derive the extinction, and assumed that it
represents accurately the extinction within our larger aperture.
The Seyfert nucleus does not
contribute importantly to the H line and MIR dust emission.  The total
H$\alpha$ flux in the central $4\arcsec \times 4\arcsec$
\citep{Ver86}, which also includes emission from adjacent ``hot spots,''
is only 8\% of that in 40\arcsec.
MIR diagnostic line ratios suggest that star formation activity dominates the
low excitation ($\leq 50$~eV) line spectrum at these wavelengths as
well as the MIR and far-infrared continuum luminosities \citep{Stu02}.
The nucleus is unresolved in the ISOCAM maps, preventing an accurate estimate
of its contribution to the MIR fluxes, but the ISOCAM CVF data do not
provide evidence for a significant AGN contribution
based on the diagnostics of \citet{Rig99} and \citet{Lau00}.
The [\ion{N}{ii}]/(H$\alpha$ $+$ [\ion{N}{ii}]) ratio in the central
$4\arcsec \times 4\arcsec$ and in several hot spots within
$14\arcsec \times 20\arcsec$ is $\approx 0.3$ \citep{All81, Ver86}.

{\em NGC\,4102} --
The nucleus is saturated in both ISOCAM LW2 and LW3 observations,
more severely for LW2. This galaxy generates a powerful outflow
detected in the Pa$\beta$ and Br$\gamma$ lines \citep{Rou03}.

{\em NGC\,4293} --
Since the central MIR source is very small compared to the pixel size of
the ISOCAM maps \citep{Rou01a}, we do not use the fitted aperture for
area normalization, but instead the size derived from the Pa$\alpha$
and H$\alpha$ maps. The data of \citet{Ver86} in the central
$4\arcsec \times 4\arcsec$ give a high line ratio
[\ion{N}{ii}]\,$\lambda\,6583$\,\AA/(H$\alpha$ +
[\ion{N}{ii}]\,$\lambda\,6583$\,\AA) $= 0.68$ due to the
Liner nucleus, and an H$\alpha$ flux accounting for $\approx 9$\%
of the flux in $d = 32.4\arcsec$.

{\em NGC\,4691} --
We combined the Br$\gamma$ measurement of \citet{Pux90} with the H$\alpha$ flux
integrated within the same region from an H$\alpha$ $+$ [\ion{N}{ii}] map 
to derive the extinction, and assumed that it remains the same
within our larger aperture.
The [\ion{N}{ii}] / (H$\alpha$ + [\ion{N}{ii}]) ratio in the central
hot spots is $\approx 0.3$ \citep{Kee83, Gar95, Gar99}.
As the central structure contains multiple knots which are partly
blended in the ISOCAM maps, and is not well represented by a single
gaussian \citep{Rou01a}, we strayed from our general definition to
determine the area normalization.
As the MIR and H$\alpha$ emission of NGC\,4691 lacks in the disk and
is very diffuse outside the central star-forming knots, we simply selected
pixels above the $3 \sigma$ brightness level in the H$\alpha$ map and
added their areas to compute an equivalent diameter.

{\em NGC\,5194} --
We considered the central emission plateau of diameter $90^{\prime\prime}$.
The motivation for this choice
instead of selecting a smaller region around the nucleus was that the plateau
represents a transition between disks and more active circumnuclear
regions in terms of MIR properties.
Besides, the emission from the Seyfert nucleus is diluted and completely
negligible within this large aperture.
We adopted the average extinction, weighted by intrinsic H$\alpha$ luminosities,
derived by \citet{Sco01} from H$\alpha$/Pa$\alpha$ decrements
of a large sample of \ion{H}{ii} regions, assuming it is representative
of the effective extinction throughout the central plateau.
We applied a small correction to this extinction to account for the
different line emissivities and extinction laws adopted.
The [\ion{N}{ii}] / (H$\alpha$ + [\ion{N}{ii}]) ratio in the central
$d \la 20\arcsec$ is high and reaches $\approx 0.85$ due to the
Seyfert nucleus but goes down to $\approx 0.3$ outside these regions
\citep{Ros82}.

{\em NGC\,5236} --
\citet{Gen98} report in the SWS $14\arcsec \times 20\arcsec$
aperture an extinction of 5~mag, larger than the value adopted here.
The ratio [\ion{N}{ii}]\,$\lambda\,6583$\,\AA/(H$\alpha$ +
[\ion{N}{ii}]\,$\lambda\,6583$\,\AA) $\approx 0.3$ in the
central $\approx 5\arcsec$ \citep{Kee84, Ver86}. The nucleus is saturated
in the ISOCAM LW2 and LW3 maps, more severely for LW3.
We thus used maps simulated from the CVF spectral cube to measure the
LW2 and LW3 fluxes. We estimate in this way that the missing flux fractions
due to saturation are 11\% and 34\% respectively, assuming that the
effects discussed in Sect.~\ref{Sub-diagn_PAH_VSG}, making photometry
from CVFs uncertain by 10--20\%, are negligible.

{\em NGC\,6946} --
We combined the Br$\gamma$ flux of \citet{Pux88} with the H$\alpha$ flux
integrated in the same aperture from an H$\alpha$ $+$ [\ion{N}{ii}] map.
\citet{Kee84} gives [\ion{N}{ii}]\,$\lambda\,6583$\,\AA/(H$\alpha$ +
[\ion{N}{ii}]\,$\lambda\,6583$\,\AA) $\approx 0.37$ in the
central 8.1\arcsec.
We assumed the same ratio throughout our larger aperture to calibrate
the H$\alpha$ $+$ [\ion{N}{ii}] map.
The nucleus is saturated in the ISOCAM LW2 and LW3 maps, more severely for LW2.
As for NGC\,5236, we used maps simulated from the CVF spectral cube, and
we estimate that the missing flux fractions due to saturation are
27\% and 13\%, respectively.

{\em NGC\,7552} --
\citet{Ver03} published Br$\alpha$ and Br$\beta$ fluxes obtained with SWS in a
$14\arcsec \times 20\arcsec$ aperture which matches fairly well
our circular aperture of 21.8\arcsec.
We used Br$\alpha$ only because of possible blending of
Br$\beta$ with H$_2$ $1 - 0~O(2)$, and combined it with the
H$\alpha$ flux integrated over the same region from an
H$\alpha$ $+$ [\ion{N}{ii}] map to derive the extinction.
The nucleus is slightly saturated in the LW3 map.
From the data of \citet{Ver86} for the central $4\arcsec \times 4\arcsec$,
[\ion{N}{ii}]\,$\lambda$\,6583\,\AA/(H$\alpha$ +
[\ion{N}{ii}]\,$\lambda$\,6583\,\AA) $= 0.37$ and the
H$\alpha$ flux is 24\% of the flux in 21.8\arcsec.

{\em NGC\,7771} --
Br$\gamma$ and radio imaging shows that the central 
$d \approx 10\arcsec$ area hosts the active star-forming regions,
distributed mainly along a circumnuclear ring \citep{Nef92, Reu00}.
Note that although \citet{Reu00} mention that they corrected for
velocity shifting of the Br$\gamma$ line outside the narrow-band
filter they used, the data from \citet{Dale03}, which we adopted,
yield a Br$\gamma$ flux almost three times higher in the central
$5\arcsec \times 10\arcsec$.

{\em NGC\,253} --
The Seyfert nature of the nucleus \citep{Ver01} is unconfirmed by other optical
spectroscopic studies, and by near- and mid-infrared
spectroscopy \citep[e.g.][]{Eng98, Stu00}; neither by our ISOCAM data
based on the diagnostics of \citet{Rig99} and \citet{Lau00}.
We used the Br$\gamma$ flux of \citet{Eng98} integrated over
$d = 15\arcsec$, which contains all the flux and coincides with
the fitted size of the MIR emission. We used their Pa$\beta$/Br$\gamma$
flux ratio measured in $2.4\arcsec \times 12\arcsec$ to derive the
extinction. Published extinction and $Q_\mathrm{Lyc}$ estimates
from H line data vary greatly.  For instance, \citet{Ver03} derived from SWS
line observations $A_V^\mathrm{MIX} \sim 9$~mag, considerably lower
than $A_V^\mathrm{MIX} = 30$~mag reported by \citet{Gen98},
although the $Q_\mathrm{Lyc}$ values are similar.

{\em NGC\,520} --
Br$\gamma$ line emission in this interacting system is detected at the
primary nucleus within a region of $\sim 5\arcsec \times 3\arcsec$
\citep{Sta91, Kot01}.  We combined the Br$\gamma$ flux of \citet{Sta91}
in the central $6\arcsec \times 8\arcsec$ with the H$\alpha$
flux measured within the same aperture from an H$\alpha$+[\ion{N}{ii}] map,
to derive the extinction. We made use of the
H$\alpha$/(H$\alpha$+[\ion{N}{ii}]) ratio of 0.58 found by \citet{Vei95}
at the nucleus. Since the primary nucleus suffers from very large extinction,
the use of H$\alpha$ photometry in a larger aperture is especially uncertain.

{\em NGC\,1808} --
We used the
Br$\gamma$ flux of \citet{Kra94} integrated in a 20\arcsec\ aperture
centered at the nucleus, which contains the quasi-totality of the flux
and coincides reasonably well with the fitted MIR size.
\citet{Kra94} derived extinction values ranging from
3 to 5~mag from H$\alpha$/Br$\gamma$ decrements, in excellent agreement
with estimates from H$\beta$/H$\alpha$ decrements quoted in the same
paper, so we applied an average of 4~mag.

{\em NGC\,3034 (M\,82)} --
We used directly the extinction and $Q_\mathrm{Lyc}$ results
of \citet{FS01} for the starburst core within $d = 30\arcsec$.
These results were derived from an extensive set of H lines from optical to
radio wavelengths which are best fitted by a mixed model and with deviations
from the \citet{Dra89} extinction law, at $\lambda = 3 - 10\,\mu$m,
as found towards the Galactic Center \citep{Lut99a}.
The MIR source is elongated along the optical major axis. Computing
the quadratic mean of the major axis and minor axis widths, we
find that $2.5 \times {\rm HPBW} = 29.95$\arcsec\ at 15\,$\mu$m,
thus extremely close to the adopted aperture of 30\arcsec. The
5-8.5\,$\mu$m emission is more extended, and we would have derived
an aperture of 35\arcsec\ from it.

{\em IC\,342} --
We adopted an aperture of $17\arcsec \times 17\arcsec$
corresponding to the size of the Br$\gamma$ line map provided by
\citet{Bok97}, which encompasses the circumnuclear starburst ring.
This aperture is close to the intrinsic size of the mid-IR source
fitted on the surface brightness profiles, between 19\arcsec\ and 22\arcsec.
The areas agree to within 30\%, and the MIR fluxes to within 9\%
in the different apertures. We combined the
Br$\gamma$ flux integrated over the map with the Br$\alpha$ flux of 
\citet{Ver03} obtained in the $14\arcsec \times 20\arcsec$
SWS beam (excluding their Br$\beta$ measurement because of possible
contribution from the H$_2$ $1 - 0~O(2)$ line and Pf$\alpha$
because of larger uncertainties on the extinction law near 7\,$\mu$m).
Fits assuming a UFS and a mixed model are both well constrained,
and the derived extinctions imply nearly identical $Q_\mathrm{Lyc}$
(within 1\%).

{\em NGC\,3256} --
The MIR emission can be separated into two components of different
sizes, a core of $\approx 8$\arcsec\ superposed onto a source of
$\approx 20$\arcsec. Therefore, a single gaussian
does not provide a good fit to the surface brightness profile.
We adopt the size of the larger source as our aperture.
\citet{Kawara87} report a Br$\gamma$ flux of $14.8 \times 10^{-17}$\,W\,m$^{-2}$
in a smaller aperture of $9\arcsec \times 18\arcsec$, almost
identical to the total Br$\gamma$ flux of $15 \times 10^{-17}$\,W\,m$^{-2}$
measured by \citet{Moo94} in a field of view of
$34\arcsec \times 34\arcsec$. We derive the extinction in
the central $3.5\arcsec \times 3.5\arcsec$ from the
Pa$\beta$/Br$\gamma$ ratio of \citet{Doy94} and assume that it does not
vary inside our larger aperture.

{\em NGC\,6240} --
The MIR emission of NGC\,6240 is unresolved in the ISOCAM data.
We adopt a size of 3\arcsec\ based on the Br$\gamma$ map of \citet{Tec00}.
As a comparison, we derive from brightness profile fitting a HPBW of
the order of 2.7\arcsec, which is ill-constrained since it is smaller
than the cleaned PSF HPBW of 3\arcsec.
We adopted the Br$\gamma$ flux of \citet{Rieke85} measured in a 8.7\arcsec\
aperture, and the Pa$\beta$ flux of \citet{Sim96} measured in a slit of width
1.5\arcsec\ oriented along the major axis of the source. In view of the
Br$\gamma$ map of \citet{Tec00}, these apertures should include the
total fluxes.
The Pa$\alpha$ flux of \citet{dePoy86}, measured in a 5.5\arcsec\ aperture,
yields a negative extinction when combined with the Pa$\beta$ flux;
it may thus be strongly underestimated.
MIR diagnostics indicate that starburst activity dominates the dust
emission and the low-excitation fine-structure line emission at these
wavelengths (\citealt{Gen98, Rig99, Lau00}; but see also \citealt{Lutz03}).

{\em IRAS\,23128-5919} --
The two galaxies in this merging pair are separated by a projected distance
of less than 5\arcsec, and are marginally resolved by ISOCAM.
We derive HPBWs of the order of 2.5\arcsec\ for both nuclei.
For lack of adequate high-resolution data, and in analogy with NGC\,6240,
we adopt a total size of 3\arcsec\ for each nucleus, thus an equivalent
size of 4.2\arcsec.
To derive the extinction, we combined measurements of pure H$\alpha$
and Pa$\alpha$ fluxes in the southern nucleus, which is much brighter
than the northern nucleus. Note that because of the narrow slit used
by \citet{Duc97}, the H$\alpha$ fluxes may be underestimated.
In the MIR, we performed the brightness profile
fitting and aperture photometry of the two blended nuclei together.
Since the CVF spectrum has poor signal to noise ratio, and the 12-18\,$\mu$m
flux simulated from the CVF is not in agreement with the observed broadband flux,
we used only broadband images. From both optical and MIR diagnostics,
this system is classified as a starburst in which the large velocities
observed in emission lines of the southern nucleus are caused by starburst
superwinds \citep{Johansson88, Lut99b}.

{\em Arp\,220} --
The two nuclei of this merger are separated by $\approx 1$\arcsec\
and are totally unresolved in the ISOCAM data. The size that we derive
from 15\,$\mu$m brightness profile fitting is 1.8-1.9\arcsec\ HPBW.
We use a total extent of 2\arcsec, as implied by the high angular
resolution observations of the mid-IR emission by \citet{Soi99}.
MIR diagnostics indicate that starburst activity dominates the dust
emission and the low-excitation fine-structure line emission
(\citealt{Gen98, Rig99, Lau00}; see also the detailed study
by \citealt{Spoon03}).
The extinction that we derive from the Br$\alpha$ and Br$\gamma$ fluxes
is in good agreement with that predicted by \citet{Ana00} from a set of
radio recombination lines.

\section{Spectral observations of IC\,342}   \label{App-IC342}

\begin{figure}[!t]
\hspace*{-1cm}
\resizebox{10cm}{!}{\rotatebox{90}{\includegraphics{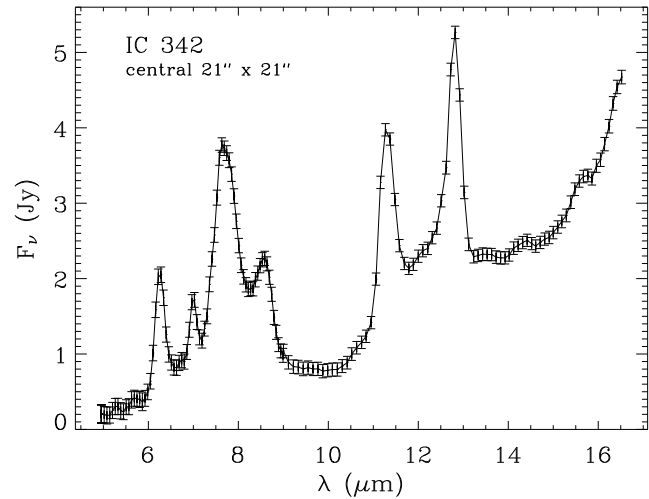}}}
\caption{
Mid-infrared spectrum of IC\,342 summed over the central $21\arcsec \times 21\arcsec$.
The error bars represent the uncertainty on the subtracted zodiacal foreground.
Errors arising from imperfect signal stabilization are not included, but are
likely important at $\lambda \geq 15.5\,\mu$m.
}
\label{fig:cvf_ic342}
\end{figure}

The CVF observations of IC\,342 consist of a single pointing centered
on the nucleus, with the  $3\arcsec$\,pixel$^{-1}$ scale giving a
$96\arcsec \times 96\arcsec$ field of view, and contain
$\approx 35$ exposures of 2.1~s per wavelength channel for a total
integration time of $\mathrm{3^{h}\,18^{m}}$.
A representative spectrum, in a region of $21\arcsec \times 21\arcsec$,
is shown in Fig.~\ref{fig:cvf_ic342}. It shows all the usual features
of star-forming galaxies, with aromatic bands nearly invariant in shape
and relative intensity (see Fig.~\ref{fig-bandpasses}). Notice the faint
emission bands that are also present in the spectra of the other galaxies of
this sample, and which are attributable to carriers of the same nature as those
emitting the bright bands at 6.2, 7.7, 8.6, 11.3 and 12.7\,$\mu$m.
Those contributing to the emission within the LW3 filter and observed
in this spectrum are detected at 12.0, 13.6, 14.3 and 15.7\,$\mu$m
\citep{Stu00}.

\bibliographystyle{apj}

\begin{thebibliography}{}

\bibitem[\protect\citeauthoryear{Alloin et al.}{1981}]{All81}
         Alloin, D., Edmunds, M. G., Lindblad, P. O., \& Pagel, B. E. J.
         1981, \aap, 101, 377
\bibitem[\protect\citeauthoryear{Anantharamaiah et al.}{2000}]{Ana00}
         Anantharamaiah, K. R., Viallefond, F., Mohan, N. R., Goss, W. M., \& Zhao, J. H.
	 2000, \apj, 537, 613
\bibitem[\protect\citeauthoryear{Biviano et al.}{1998a}]{Biv98a}
         Biviano, A., Altieri, B., Blommaert, J., et al. 1998a,
         The ISOCAM CVF Calibration Report, version 1.1,
         ESA/CAM IDT Vilspa Technical report
\bibitem[\protect\citeauthoryear{Biviano et al.}{1998b}]{Biv98b}
         Biviano, A., Blommaert, J., Laurent, O., et al. 1998b,
         The ISOCAM Flat Field Calibration Report, version 1.1,
	 ESA/CAM IDT Vilspa Technical report
\bibitem[\protect\citeauthoryear{B\"oker et al.}{1999}]{Bok99}
         B\"oker, T., Calzetti, D., Sparks, W., et al. 1999, \apjs, 124, 95
\bibitem[\protect\citeauthoryear{B\"oker et al.}{1997}]{Bok97}
         B\"oker, T., F\"orster Schreiber, N. M., \& Genzel, R. 1997, \aj, 114, 1883
\bibitem[\protect\citeauthoryear{Boselli et al.}{1998}]{Bos98}
         Boselli, A., Lequeux, J., Sauvage, M., et al. 1998, \aap, 335, 53
\bibitem[\protect\citeauthoryear{Boulanger et al.}{1998}]{Boulanger98}
         Boulanger, F., Abergel, A., Bernard, J. P. et al. 1998,
         ASP Conference Series, 132, 15
\bibitem[\protect\citeauthoryear{Cardelli et al.}{1989}]{Car89}
         Cardelli, J. A., Clayton, G. C., \& Mathis, J. S. 1989, \apj, 345, 245
\bibitem[\protect\citeauthoryear{Cesarsky et al.}{1996}]{Ces96}
         Cesarsky, C. J., Abergel, A., Agn\`ese, P., et al. 1996, \aap, 315, L32
\bibitem[\protect\citeauthoryear{Cesarsky \& Sauvage}{1999}]{Ces99}
         Cesarsky, C. J., \& Sauvage, M. 1999, \apss, 269, 303
\bibitem[\protect\citeauthoryear{Charmandaris et al.}{2002}]{Cha02}
         Charmandaris, V., Laurent, O., Le Floc'h, E., et al.~2002, \aap, 391, 429
\bibitem[\protect\citeauthoryear{Contursi et al.}{2000}]{Con00}
         Contursi, A., Lequeux, J., Cesarsky, C., et al. 2000, \aap, 362, 310
\bibitem[\protect\citeauthoryear{Coulais \& Abergel}{2000}]{Coulais00}
         Coulais, A., \& Abergel, A. 2000, \aaps, 141, 533
\bibitem[\protect\citeauthoryear{Cr\'et\'e et al.}{1999}]{Cre99}
         Cr\'et\'e, E., Giard, M., Joblin, C., et al. 1999, \aap, 352, 277
\bibitem[\protect\citeauthoryear{Dale et al.}{2001}]{Dal01}
         Dale, D. A., Helou, G., Contursi, A., Silbermann, N. A., \& Kolhatkar, S.
         2001, \apj, 549, 215
\bibitem[\protect\citeauthoryear{Dale \& Helou}{2002}]{Dale02}
         Dale, D. A., \& Helou, G. 2002, \apj, 576, 159
\bibitem[\protect\citeauthoryear{Dale et al.}{2003}]{Dale03}
         Dale, D. A., Roussel, H., Contursi A. et al., 2003, \apj, in press
         (astro-ph/0310383)
\bibitem[\protect\citeauthoryear{Dale et al.}{2000}]{Dal00}
         Dale, D. A., Silbermann, N. A., Helou, G., et al. 2000, \aj, 120, 583
\bibitem[\protect\citeauthoryear{Davidge \& Pritchet}{1990}]{Dav90}
         Davidge, T. G., \& Pritchet, C. J. 1990, \aj, 100, 102
\bibitem[\protect\citeauthoryear{de Graauw et al.}{1996}]{deG96}
         de Graauw, T., Haser, L. N., Beintema, D. A., et al. 1996, \aap, 315, L49
\bibitem[\protect\citeauthoryear{de Poy et al.}{1986}]{dePoy86}
         de Poy, D. L., Becklin, E. E. \& Wynn-Williams, C. G. 1986, \apj, 307, 116
\bibitem[\protect\citeauthoryear{D\'esert et al.}{1990}]{Des90}
         D\'esert, F. X., Boulanger, F., \& Puget, J. L. 1990, \aap, 237, 215
\bibitem[\protect\citeauthoryear{de Vaucouleurs et al.}{1991}]{deV91}
         de Vaucouleurs, G., de Vaucouleurs, A.,  Corwin, H. G., et al. 1991,
         Third Reference Catalog of Bright Galaxies (RC3)
\bibitem[\protect\citeauthoryear{Devereux \& Young}{1993}]{Dev93}
         Devereux, N. A., \& Young, J. S. 1993, \aj, 106, 948
\bibitem[\protect\citeauthoryear{Doyon et al.}{1994}]{Doy94}
         Doyon, R., Joseph, R. D., \& Wright, G. S. 1994, \apj, 421, 101
\bibitem[\protect\citeauthoryear{Draine}{1989}]{Dra89}
         Draine, B. T. 1989, in Proc. of the $\rm 22^{nd}$ ESLAB Symp.,
         Infrared Spectroscopy in Astronomy, ed. B. H. Kaldeich (ESA SP-290), 93
\bibitem[\protect\citeauthoryear{Draine \& Anderson}{1985}]{Draine85}
         Draine, B. T. \& Anderson, N. 1985, \apj, 292, 494
\bibitem[\protect\citeauthoryear{Duc et al.}{1997}]{Duc97}
         Duc, P.-A., Mirabel, I. F., \& Maza, J. 1997, \aaps, 124, 533
\bibitem[\protect\citeauthoryear{Engelbracht et al.}{1998}]{Eng98}
         Engelbracht, C. W., Rieke, M. J., Rieke, G. H., Kelly, D. M., \& Achtermann, J. M.
         1998, \apj, 505, 639
\bibitem[\protect\citeauthoryear{F\"orster Schreiber et al.}{2001}]{FS01}
         F\"orster Schreiber, N. M., Genzel, R., Lutz, D., Kunze, D., \& Sternberg, A.
         2001, \apj, 552, 544
\bibitem[\protect\citeauthoryear{F\"orster Schreiber et al.}{2003}]{FS03}
         F\"orster Schreiber, N. M., Sauvage, M., Charmandaris, V., et al.
	 2003, \aap, 399, 833
\bibitem[\protect\citeauthoryear{Freedman \& Madore}{1988}]{Fre88}
         Freedman, W. L., \& Madore, B. F. 1988, \apj, 332, L63
\bibitem[\protect\citeauthoryear{Garc\'{\i}a-Barreto et al.}{1999}]{Gar99}
         Garc\'{\i}a-Barreto, J. A., Aceves, H., Kuhn, O, et al.
         1999, Rev. Mex. Astron. Astrofis. 35, 173
\bibitem[\protect\citeauthoryear{Garc\'{\i}a-Barreto et al.}{1995}]{Gar95}
         Garc\'{\i}a-Barreto, J. A., Franco, J., Guichard, J., \& Carrillo, R.
         1995, \apj, 451, 156
\bibitem[\protect\citeauthoryear{Geballe}{1997}]{Geb97}
         Geballe, T. R. 1997, in ASP Conf. Ser. 122, From Stardust to Planetesimals,
         ed. Y. J. Pendleton \& A. G. G. M. Tielens (San Francisco: ASP), 119
\bibitem[\protect\citeauthoryear{Genzel \& Cesarsky}{2000}]{Gen00}
         Genzel, R., \& Cesarsky, C. J. 2000, \araa, 38, 761
\bibitem[\protect\citeauthoryear{Genzel et al.}{1998}]{Gen98}
         Genzel, R., Lutz, D., Strum, E., et al. 1998, \apj, 498, 579
\bibitem[\protect\citeauthoryear{Giard et al.}{1994}]{Giard94}
         Giard, M., Bernard, J. P., Lacombe, F., Normand, P., \& Rouan, D.
         1994, \aap, 291, 239
\bibitem[\protect\citeauthoryear{Giveon et al.}{2002}]{Giv02}
         Giveon, U., Sternberg, A., Lutz, D., Feuchtgruber, H., \& Pauldrach, A. W. A.
         2002, \apj, 566, 880
\bibitem[\protect\citeauthoryear{Goldader et al.}{1995}]{Gol95}
         Goldader, J. D., Joseph, R. D., Doyon, R., \& Sanders, D. B. 1995, \apj, 444, 97
\bibitem[\protect\citeauthoryear{Greenawalt et al.}{1998}]{Gre98}
         Greenawalt, B., Walterbos, R. A. M., Thilker, D., et al. 1998, \apj, 506, 135
\bibitem[\protect\citeauthoryear{Greenberg \& Hong}{1985}]{Greenberg74}
         Greenberg, J. M. \& Hong, S.-S. 1974, IAUS, 60, 155
\bibitem[\protect\citeauthoryear{Hameed \& Devereux}{1999}]{Ham99}
         Hameed, S., \& Devereux, N. A. 1999, \aj, 118, 730
\bibitem[\protect\citeauthoryear{Helou}{1986}]{Hel86}
         Helou, G. 1986, \apj, 311, L33
\bibitem[\protect\citeauthoryear{Helou et al.}{2000}]{Hel00}
         Helou, G., Lu, N. Y., Werner, M. W., Malhotra, A., \& Silbermann, N.
         2000, \apj, 532, L21
\bibitem[\protect\citeauthoryear{Hibbard \& van Gorkom}{1996}]{Hibbard96}
         Hibbard, J. E. \& van Gorkom, J. H. 1996, \aj, 111, 655
\bibitem[\protect\citeauthoryear{Hony et al.}{2001}]{Hony01}
         Hony, S., Van Kerckhoven, C., Peeters, E., Tielens, A. G., Hudgins, D. M.
         \& Allamandola, L. J. 2001, \aap, 370, 1030
\bibitem[\protect\citeauthoryear{Hummel et al.}{1987}]{Hum87}
         Hummel, E., van der Hulst, J. M., \& Keel, W. C. 1987, \aap, 172, 32
\bibitem[\protect\citeauthoryear{Jansen et al.}{2000}]{Jansen00}
         Jansen, R.A., Fabricant, D., Franx, M., \& Caldwell, N. 2000, \apjs, 126, 331
\bibitem[\protect\citeauthoryear{Johansson \& Bergvall}{1988}]{Johansson88}
         Johansson, L. \& Bergvall, N. 1988, \aap, 192, 81
\bibitem[\protect\citeauthoryear{Kawara et al.}{1987}]{Kawara87} 
         Kawara, K., Nishida, M., \& Gregory, B. 1987, \apj, 321, L35
\bibitem[\protect\citeauthoryear{Keel}{1983}]{Kee83}
         Keel, W. C. 1983, \apjs, 52, 229
\bibitem[\protect\citeauthoryear{Keel}{1984}]{Kee84}
         Keel, W. C. 1984, \apj, 282, 75
\bibitem[\protect\citeauthoryear{Kennicutt}{1983}]{Ken83}
         Kennicutt, R. C., Jr. 1983, \apj, 272, 54
\bibitem[\protect\citeauthoryear{Kennicutt}{1992}]{Kennicutt92}
         Kennicutt, R. C., Jr. 1992, \apj, 388, 310
\bibitem[\protect\citeauthoryear{Kennicutt}{1998}]{Ken98}
         Kennicutt, R. C., Jr. 1998, \araa, 36, 189 
\bibitem[\protect\citeauthoryear{Kennicutt et al.}{1989}]{Ken89}
         Kennicutt, R. C., Jr., Keel, W. C., \& Blaha, C. A. 1989, \aj, 97, 1022
\bibitem[\protect\citeauthoryear{Kessler et al.}{1996}]{Kes96}
         Kessler, M. F., Steinz, J. A., Anderegg, M. E., et al.~1996, \aap, 315, L27
\bibitem[\protect\citeauthoryear{Kewley et al.}{2001}]{Kewley01}
         Kewley, L. J., Heisler, C. A., Dopita, M. A., \& Lumsden, S. 2001, \apjs, 132, 37
\bibitem[\protect\citeauthoryear{Kim et al.}{1998}]{Kim98}
         Kim, D.-C., Veilleux, S., \& Sanders, D. B. 1998, \apj, 508, 627
\bibitem[\protect\citeauthoryear{Koopmann et al.}{2001}]{Koo01}
         Koopmann, R. A., Kenney, J. D. P., \& Young, J. S. 2001, \apjs, 135, 125
\bibitem[\protect\citeauthoryear{Kotilainen et al.}{2000}]{Kot00}
         Kotilainen, J. K., Reunanen, J., Laine, S., \& Ryder, S. D. 2000, \aap, 353, 834
\bibitem[\protect\citeauthoryear{Kotilainen et al.}{2001}]{Kot01}
         Kotilainen, J. K., Reunanen, J., Laine, S., \& Ryder, S. D. 2001, \aap, 366, 439
\bibitem[\protect\citeauthoryear{Krabbe et al.}{1994}]{Kra94}
         Krabbe, A., Sternberg, A., \& Genzel, R. 1994, \apj, 425, 72
\bibitem[\protect\citeauthoryear{Kristen et al.}{1997}]{Kri97}
         Kristen, H., J\"ors\"ater, S., Lindblad, P. O., \& Boksenberg, A. 1997, \aap, 328, 483
\bibitem[\protect\citeauthoryear{Larsen \& Richtler}{1999}]{Lar99}
         Larsen, S. S., \& Richtler, T. 1999, \aap, 345, 59
\bibitem[\protect\citeauthoryear{Laurent}{1999}]{Lau99}
         Laurent, O., 1999, Ph.D. Thesis, Universit\'e de Paris XI
\bibitem[\protect\citeauthoryear{Laurent et al.}{2000}]{Lau00}
         Laurent, O., Mirabel, I. F., Charmandaris, V., et al. 2000, \aap, 359, 887 
\bibitem[\protect\citeauthoryear{L\'eger \& Puget}{1984}]{Leg84}
         L\'eger, A., \& Puget, J. L. 1984, \aap, 135, L5
\bibitem[\protect\citeauthoryear{Li \& Draine}{2002}]{Li02}
         Li, A., \& Draine, B. T. 2002, \apj, 572, 232
\bibitem[\protect\citeauthoryear{Lindblad}{1999}]{Lin99}
         Lindblad, P. O. 1999, \aapr, 9, 221
\bibitem[\protect\citeauthoryear{Lu et al.}{2003}]{Lu03}
         Lu, N., Helou, G., Werner, M.W., et al. 2003, \apj, 588, 199
\bibitem[\protect\citeauthoryear{Lutz}{1999a}]{Lut99a}
         Lutz, D. 1999a, in The Universe as seen by {\em ISO\/},
         ed. P. Cox \& M.  F. Kessler (ESA SP-427; Noordwijk: ESA), 623
\bibitem[\protect\citeauthoryear{Lutz et al.}{2003}]{Lutz03}
         Lutz, D., Sturm, E., Genzel, R., et al. 2003, \aap, 409, 867
\bibitem[\protect\citeauthoryear{Lutz et al.}{1999b}]{Lut99b}
         Lutz, D., Veilleux, S., \& Genzel, R. 1999b, \apj, 517, L13
\bibitem[\protect\citeauthoryear{Malhotra et al.}{1996}]{Mal96}
         Malhotra, S., Helou, G., van Buren, D., et al. 1996, \aap, 315, L161
\bibitem[\protect\citeauthoryear{Mathis et al.}{1983}]{Mathis83}
         Mathis, J. S., Mezger, P. G., \& Panagia, N. 1983, \aap, 128, 212
\bibitem[\protect\citeauthoryear{Mattila et al.}{1999}]{Mat99}
         Mattila, K., Lehtinen, K., \& Lemke, D. 1999, \aap, 342, 643
\bibitem[\protect\citeauthoryear{Mattila et al.}{1996}]{Mat96}
         Mattila, K., Lemke, D., Haikala, L. K., et al. 1996, \aap, 315, L353
\bibitem[\protect\citeauthoryear{Mennella et al.}{1997}]{Mennella97}
         Mennella, V., Baratta, G. A., Colangeli, L., et al. 1997, \apj, 481, 545
\bibitem[\protect\citeauthoryear{Mirabel et al.}{1998}]{Mir98}
         Mirabel, I. F., Vigroux, L., Charmandaris, V., et al. 1998, \aap, 333, L1
\bibitem[\protect\citeauthoryear{Moorwood \& Oliva}{1994}]{Moo94}
         Moorwood, A. F. M., \& Oliva, E. 1994, \apj, 429, 602
\bibitem[\protect\citeauthoryear{Neff \& Hutchings}{1992}]{Nef92}
         Neff, S. G., \& Hutchings, J. B. 1992, \aj, 103, 1746
\bibitem[\protect\citeauthoryear{Normand et al.}{1995}]{Normand95}
         Normand, P., Rouan, D., Lacombe, F., \& Tiph\`ene, D. 1995, \aap, 297, 311
\bibitem[\protect\citeauthoryear{Okumura}{2000}]{Oku00}
         Okumura, K. 2000, in {\em ISO\/} Beyond Point Sources: Studies of Extended
         Infrared Emission, ed. R. J. Laureijs, K. Leech, \& M. F. Kessler, ESA SP-455, 47
\bibitem[\protect\citeauthoryear{Peeters et al.}{2002}]{Peeters02}
         Peeters, E., Hony, S., van Kerckhoven, C., et al. 2002, \aap, 390, 1089
\bibitem[\protect\citeauthoryear{Puget \& L\'eger}{1989}]{Puget89}
         Puget, J.L., \& L\'eger, A. 1989, \araa, 27, 161
\bibitem[\protect\citeauthoryear{Puxley et al.}{1988}]{Pux88}
         Puxley, P. J., Hawarden, T. G., \& Mountain, C. M. 1988, \mnras, 234, SC29
\bibitem[\protect\citeauthoryear{Puxley et al.}{1990}]{Pux90}
         Puxley, P. J., Hawarden, T. G., \& Mountain, C. M. 1990, \apj, 364, 77
\bibitem[\protect\citeauthoryear{Reunanen et al.}{2000}]{Reu00}
         Reunanen, J., Kotilainen, J. K., Laine, S., \& Ryder, S. D. 2000, \apj, 529, 853
\bibitem[\protect\citeauthoryear{Rieke et al.}{1985}]{Rieke85}
         Rieke, G. H., Cutri, R. M., Black, J. H. et al. 1985, \apj, 290, 116
\bibitem[\protect\citeauthoryear{Rigopoulou et al.}{1999}]{Rig99}
         Rigopoulou, D., Spoon, H. W. W., Genzel, R., et al. 1999, \aj, 118, 2625
\bibitem[\protect\citeauthoryear{Rose \& Searle}{1982}]{Ros82}
         Rose, J. A. \& Searle, L. 1982, \apj, 253, 556
\bibitem[\protect\citeauthoryear{Roussel et al.}{2003}]{Rou03}
         Roussel, H., Helou, G., Beck, R., Condon, J. J., Bosma, A.,
         Matthews, K., \& Jarrett, T. H. 2003, \apj, 593, 733
\bibitem[\protect\citeauthoryear{Roussel et al.}{2001a}]{Rou01a}
         Roussel, H., Vigroux, L., Bosma, A., et al. 2001a, \aap, 369, 473
\bibitem[\protect\citeauthoryear{Roussel et al.}{2001b}]{Rou01b}
         Roussel, H., Sauvage, M., Vigroux, L., et al. 2001b, \aap, 372, 406
\bibitem[\protect\citeauthoryear{Roussel et al.}{2001c}]{Rou01c}
         Roussel, H., Sauvage, M., Vigroux, L., \& Bosma, A. 2001c, \aap, 372, 427
\bibitem[\protect\citeauthoryear{Ryder et al.}{1995}]{Ryd95}
         Ryder, S. D., Hungerford, A., Dopita, M. A., et al. 1995,
         in The Opacity of Spiral Disks, eds. J. I. Davies \& D. Burstein (Dordrecht: Kluwer), 359
\bibitem[\protect\citeauthoryear{Saha et al.}{2002}]{Sah02}
         Saha, A., Claver, J., \& Hoessel, J. G. 2002, \aj, 124, 839
\bibitem[\protect\citeauthoryear{Sanders \& Mirabel}{1996}]{San96}
         Sanders, D. B., \& Mirabel, I. F. 1996, \araa, 34, 749
\bibitem[\protect\citeauthoryear{Sauvage et al.}{1996}]{Sau96}
         Sauvage, M., Blommaert, J., Boulanger, F., et al. 1996, \aap, 315, L89
\bibitem[\protect\citeauthoryear{Sauvage \& Thuan}{1992}]{Sau92}
         Sauvage, M., \& Thuan, T.X. 1992, \apj, 396, L69
\bibitem[\protect\citeauthoryear{Scoville et al.}{2001}]{Sco01}
         Scoville, N. Z., Polletta, M., Ewald, S., et al. 2001, \aj, 122, 3017
\bibitem[\protect\citeauthoryear{Sellgren et al.}{1990}]{Sel90}
         Sellgren, K., Luan, L., \& Werner, M. W. 1990, \apj, 359, 384
\bibitem[\protect\citeauthoryear{Shaver et al.}{1983}]{Sha83}
         Shaver, P. A., McGee, R. X., Newton, L. M., Danks, A. C., \& Pottasch, S. R.
         1983, \mnras, 204, 53
\bibitem[\protect\citeauthoryear{Simpson et al.}{1996}]{Sim96}
         Simpson, C., Forbes, D. A., Baker, A. C., \& Ward, M. J. 1996, \mnras, 283, 777
\bibitem[\protect\citeauthoryear{Smith}{1975}]{Smi75}
         Smith, H. E. 1975, \apj, 199, 591
\bibitem[\protect\citeauthoryear{Soifer et al.}{1999}]{Soi99}
         Soifer, B. T., Neugebauer, G., Matthews, K., et al. 1999, \apj, 513, 207
\bibitem[\protect\citeauthoryear{Spoon et al.}{2003}]{Spoon03}
         Spoon, H. W., Moorwood, A. F., Lutz, D., et al. 2003, astro-ph/0310721
\bibitem[\protect\citeauthoryear{Stanford}{1991}]{Sta91}
         Stanford, S. A. 1991, \apj, 381, 409
\bibitem[\protect\citeauthoryear{Storchi-Bergman et al.}{1996}]{Sto96}
         Storchi-Bergman, T., Wilson, A. S., \& Baldwin, J. A. 1996, \apj, 460, 252
\bibitem[\protect\citeauthoryear{Storey \& Hummer}{1995}]{Sto95}
         Storey, P. J., \& Hummer, D. G. 1995, \mnras, 272, 41
\bibitem[\protect\citeauthoryear{Sturm et al.}{1996}]{Sturm96}
         Sturm, E., Lutz, D., Genzel, R. et al. 1996, \aap, 315, L133
\bibitem[\protect\citeauthoryear{Sturm et al.}{2000}]{Stu00}
         Sturm, E., Lutz, D., Tran, D., et al. 2000, \aap, 358, 481
\bibitem[\protect\citeauthoryear{Sturm et al.}{2002}]{Stu02}
         Sturm, E., Lutz, D., Verma, A., et al. 2002, \aap, 393, 821
\bibitem[\protect\citeauthoryear{Tecza et al.}{2000}]{Tec00}
         Tecza, M., Genzel, R., Tacconi, L. J., et al., 2000, \apj, 537, 178
\bibitem[\protect\citeauthoryear{Thornley et al.}{2000}]{Tho00}
         Thornley, M. D., F\"orster Schreiber, N. M., Lutz, D., et al. 2000, \apj, 539, 641
\bibitem[\protect\citeauthoryear{Tokunaga}{1997}]{Tok97}
         Tokunaga, A. 1997, in ASP Conf. Ser. 124, Diffuse Infrared Radiation
         and the {\em IRTS}, ed. H. Okuda, T. Matsumoto, \& T. L. Roellig (San Francisco: ASP), 149
\bibitem[\protect\citeauthoryear{Tran}{1998}]{Tran98}
         Tran Q. D., 1998, PhD thesis, Universit\'e de Paris XI
\bibitem[\protect\citeauthoryear{Tran et al.}{2001}]{Tran01}
         Tran, Q. D., Lutz, D., Genzel, R., et al. 2001, \apj, 552, 527
\bibitem[\protect\citeauthoryear{Tully}{1988}]{Tul88}
         Tully, R. B. 1988, Nearby Galaxies Catalog (Cambridge University Press)
\bibitem[\protect\citeauthoryear{Uchida et al.}{1998}]{Uch98}
         Uchida, K. I., Sellgren, K., \& Werner, M. W. 1998, \apj, 493, L108
\bibitem[\protect\citeauthoryear{Uchida et al.}{2000}]{Uch00}
         Uchida, K. I., Sellgren, K., Werner, M. W., \& Houdashelt, M. L.
         2000, \apj, 530, 817
\bibitem[\protect\citeauthoryear{Veilleux et al.}{1995}]{Vei95}
         Veilleux, S., Kim, D.-C., Sanders, D. B., Mazzarella, J. M., \& Soifer, B. T.
	 1995, \apjs, 98, 171
\bibitem[\protect\citeauthoryear{Verma et al.}{2003}]{Ver03}
         Verma, A., Lutz, D., Sturm, E., Sternberg, A., Genzel, R., \& Vacca, W.
         2003, \aap, 403, 829
\bibitem[\protect\citeauthoryear{V\'eron-Cetty \& V\'eron}{1986}]{Ver86}
         V\'eron-Cetty, M.-P., \& V\'eron, P. 1986, \aaps, 66, 335
\bibitem[\protect\citeauthoryear{V\'eron-Cetty \& V\'eron}{2001}]{Ver01}
         V\'eron-Cetty, M.-P., \& V\'eron, P. 2001, \aap, 374, 92
\bibitem[\protect\citeauthoryear{Verstraete et al.}{1996}]{Ver96}
         Verstraete, L., Puget, J.-L., Falgarone, E., et al. 1996, \aap, 315, L337

\end{thebibliography}

\end{document}